\documentclass[12pt]{article}
\usepackage{yfonts}
\DeclareMathAlphabet{\scr}{U}{rsfs}{m}{n}
\usepackage{supertabular}
\usepackage{amsmath}
\usepackage{amssymb}
\usepackage{bbold}
\usepackage[sans]{dsfont}
\usepackage{graphicx}
\usepackage{epsfig}
\usepackage{wasysym}
\usepackage[dvips]{color}

\newcommand{\cleqn}{\setcounter{equation}{0}}
\newcommand{\newc}{\newcommand}
\newc{\eps}{\epsilon}
\newc{\lam}{\lambda}
\newc{\Lam}{\Lambda}
\newc{\ra}{\rightarrow}
\newc{\wtilde}{\widetilde}
\newc{\ie}{\textit{i.e.}\;\,}
\newc{\eg}{\textit{e.g.}\;\,}
\newc{\rpv}{\not\!\! M_p}
\newc{\lsim}{\stackrel{<}{\sim}}
\newc{\beq}{\begin{equation}}
\newc{\eeq}{\end{equation}}
\newc{\beqn}{\begin{eqnarray}}
\newc{\eeqn}{\end{eqnarray}}
\newc{\PLB}{\emph{Phys.Lett.}{\bf{B}}}
\newc{\NPB}{\emph{Nucl.Phys.}{\bf{B}}}
\newc{\mcal}{\mathcal}
\newc{\bsym}{\boldsymbol}
\newc{\nonum}{\nonumber}
\newc{\ol}{\overline}
\definecolor{Red}{cmyk}{0,1,1,0}
\definecolor{luhn}{rgb}{1,0,0.5}
\definecolor{thor}{rgb}{0,0.6,1}
\definecolor{BurntOrange}{cmyk}{0,0.51,1,0}
\definecolor{BrickRed}{cmyk}{0,0.89,0.94,0.28}
\begin{document}
\setlength{\baselineskip}{.55cm}
\vspace{-1cm}
\title{\hfill ~\\[-20mm]
       \hfill\mbox{\small UFIFT-HEP-07-11}\\[8mm]
\textbf{Proton Hexality from an Anomalous \\ 
Flavor $\bsym{U\!(1)}$ and Neutrino Masses -- \\
Linking to the String Scale} } \date{}
\author{\\Herbi K. Dreiner$^1$,\footnote{ E-mail: {\tt
      dreiner@th.physik.uni-bonn.de}}~~ Christoph Luhn$^{2}$,\footnote{
    E-mail: {\tt luhn@phys.ufl.edu}}\\
  Hitoshi Murayama$^{3,4}$,\footnote{ E-mail: {\tt
      murayama@hitoshi.berkeley.edu}}~~ Marc Thormeier$^5$\footnote{
    E-mail: {\tt thor@th.physik.uni-bonn.de}}
  \\
  \\
  \emph{$^1$\small{}Physikalisches Institut der Universit\"at Bonn,}\\
  \emph{\small Nu\ss allee 12, 53115 Bonn, Germany}\\[4mm]
 \emph{$^2$\small{}Institute for Fundamental Theory, Department of Physics,}\\
  \emph{\small University of Florida, Gainesville, FL 32611, USA}\\[4mm]
  \emph{$^3$\small{}Theoretical Physics Group,}\\\emph{\small Ernest
    Orlando Lawrence     Berkeley National Laboratory,}\\
  \emph{\small University of California, Berkeley, CA 94720, USA}\\[4mm]
  \emph{$^4$\small{}Department of Physics,}\\\emph{\small University of
    California, Berkeley, CA 94720, USA}\\[4mm]
  \emph{$^5$\small{}Service de Physique Th\'eorique, CEA-Saclay,}\\
  \emph{\small Orme des Merisiers, 91191 Gif-sur-Yvette Cedex, France}}

\maketitle

\begin{abstract}
\vspace{.1cm}
\noindent
We devise minimalistic gauged $U(1)_X$ Froggatt-Nielsen models which
at low-energy give rise to the recently suggested discrete gauge
$\mathbb{Z}_6$-symmetry, proton hexality, thus stabilizing the proton.
Assuming three generations of
right-handed neutrinos, with the proper choice of $X$-charges, we
obtain viable neutrino masses. Furthermore, we find scenarios such
that no $X$-charged hidden sector superfields are needed, which from a
bottom-up perspective allows the calculation of $g_{\mathrm{string}}$,
$g_X$ and $G_{\mathrm{SM}}$'s Ka\v{c}-Moody levels. The only mass
scale apart from $M_{\mathrm{grav}}$ is $m_{\mathrm{soft}}$.

\end{abstract}





\section{Introduction}
In this paper, we consider low-energy discrete symmetries, $\mathbb
{Z}_{N}$, as extensions of the $SU(3)\times SU(2)\times U(1)$ gauge
symmetry of the Minimal Supersymmetric Standard Model (MSSM). We focus
on the case, where the $\mathbb{Z}_{N}$ is the remnant of a
spontaneously broken local gauge symmetry, in order to avoid potentially
harmful gravity effects \cite{Krauss:1988zc}. Such discrete symmetries
originating in a gauge theory are called discrete gauge symmetries
(DGSs) \cite{ Wegner:1984qt}. In Refs.~\cite{Ibanez:1991pr,
Dreiner:2005rd}, a systematic study was performed of all the DGSs
resulting from Abelian, anomaly-free gauge symmetries, $U(1)_X$, which
leave the MSSM invariant.  Specifically, the following assumptions
were made in these studies\footnote{In Ref.~\cite{Luhn:2007gq}, the
case will be investigated where these points are modified such that
massless right-handed neutrinos exist, hence the possible DGSs in
combination with Dirac rather than Majorana neutrinos will be
explored.}
\begin{itemize}
\item The only light, low-energy fields are those of the MSSM. All
  beyond-the-MSSM fields are heavy.
\item At least the following superpotential terms are $\mathbb{Z}_N
$-invariant:
\beq
Q^iH^D\overline{D^j},\;\; Q^iH^U\overline{U^j},\;\;L^iH^D\overline{E^j},\;\;
H^DH^U,\;\; L^iH^UL^jH^U\,,
\eeq
where we have made use of the standard notation for the MSSM chiral
superfields, see for example \cite{Allanach:2003eb}. The invariance of
the first three terms implies that the $\mathbb{Z}_N$-symmetry, but
not necessarily the original $U(1)_X$, is family-universal.
\end{itemize}
Given these assumptions, the only possible DGS resulting from an
anomaly-free $U(1)_X$ are the $\mathbb{Z}_2 $-symmetry matter parity
($M_p$), the $\mathbb{Z}_3$-symmetry baryon triality ($B_3$) and the
$\mathbb{Z}_6 $-symmetry proton hexality ($P_6=M_p\times B_3$)
\cite{Ibanez:1991pr,Dreiner:2005rd}. In Refs.~\cite{Dreiner:2003yr,
  Dreiner:2006xw}, the $U(1 )_X$ gauge charges were determined, which
lead to a low-energy $M_p, \;B_3$, or $P_6$, respectively. See also
Refs.~\cite{Martin:1992mq,Mohapatra:2007vd} for related work on the conditions
for DGSs in GUTs.

It is now of great interest to see whether realistic flavor models
for the Standard Model (SM) fermion masses and mixings can be
constructed in each case. Employing the original $U(1)_X$ in a minimal
Froggatt-Nielsen (FN) scenario \cite{Froggatt:1978nt} and using the
Green-Schwarz (GS) mechanism \cite{Green:1984sg} to cancel the $U(1)_X
$ anomalies, a successful $M_p$-model was constructed in 
Ref.~\cite{Dreiner:2003yr} and its implications for suppressed proton decay
were discussed in Refs.~\cite{Harnik:2004yp,Larson:2004ji}. Later, a
corresponding 
$B_3$-model was constructed in Ref.~\cite{ Dreiner:2006xw}, with a detailed
discussion of the neutrino masses. 

It is the purpose of this note to construct a $P_6$-FN flavor model,
in order to complete this program. Furthermore, from the
phenomenological point of view, proton hexality is a very attractive
symmetry. It combines the advantages of the $M_p$ and the $B_3$ models
\cite{Dreiner:2005rd}: the lightest supersymmetric particle (LSP) is
stable and the dangerous dimension-four and dimension-five proton
decay operators are forbidden. We shall proceed analogously to
Refs.~\cite{Dreiner:2003yr, Dreiner:2006xw} and refer the reader to
these publications for an explanation of our notation and an
introduction to for example the Giudice-Masiero/Kim-Nilles (GM/KN)
mechanism \cite{Giudice:1988yz, Kim:1994eu}.

There has been extensive previous work on anomalous flavor models
employing the Green-Schwarz mechanism and with breaking slightly below
the Planck scale, see for example Refs.~\cite{Ibanez:1994ig,
Jain:1994hd,Dudas:1995yu,Binetruy:1996xk,Irges:1998ax,Maekawa:2001uk}.
However, we believe this is the first work on such a model aiming for a
remnant ``gauged'' $P_6$. There are also some non-anomalous flavor
models with $U(1)_X$ breaking at the TeV scale \cite{Cvetic:1997ky,
Langacker:1998tc,Demir:2005ti,Chen:2006hn,Lee:2007fw,LeLuMa,MarcCarlos}.

This note is structured as follows: In Sect.~\ref{nonusect}, we discuss
the constraints on the $X$-charges which are not related to neutrino
phenomenology. In Sect.~\ref{nusect}, we then focus on the neutrino
sector and how it fixes the $X$-charges; corresponding tables are
given in Appendix~\ref{tablesect}. In Sect.~\ref{anomaliesetc}, we
discuss the possibility and the implications of excluding $X$-charged
hidden sector superfields, enabling us to calculate the string coupling
constant. We conclude in Sect.~\ref{conclusionsect}.





\cleqn
\section{\label{nonusect}Non-Neutrino 
Constraints on the $\boldsymbol{X}$-Charges}

In the following we proceed as in Refs.~\cite{Dreiner:2003yr,
Dreiner:2006xw} and consider only one flavon chiral superfield $A$,
with $U(1)_X$-charge $X_A=-1$. In order to obtain a viable flavor
model, the $U(1)_X$ charges of the ${P}_6$--FN models must satisfy
several phenomenological and consistency constraints. They must
\begin{itemize}
\item[($a$)] reproduce phenomenologically acceptable charged SM
  fermion masses and mixings, see Ref.~\cite{Ross:2007az},
\item[($b$)] reproduce phenomenologically acceptable neutrino masses
  and mixings,
\item[($c$)] satisfy the Green-Schwarz mixed linear anomaly
  cancellation conditions (with gauge coupling unification), as well
  as guarantee that the mixed quadratic anomaly vanishes on its
  own, \eg Ref.~\cite{Dreiner:2003yr},
\item[($d$)] imply the desired low-energy DGS ${P}_6$, \ie
  give rise to the following discrete family-independent
  $\mathbb{Z}_6$-charges for the MSSM chiral superfields
  \cite{Dreiner:2005rd,Babu:2003qh}: 
  $$
  z_Q=0, ~~z_{\overline{D}}=5,
  ~~z_{\overline{U}}=1, ~~z_L=4, ~~z_{\overline{E}}=1, ~~z_{H^D}=1,
  ~~z_{H^U}=5\,,
  $$ 
  and (as will be argued later) $z_{\overline{N}}=3$
  for the additional right-handed neutrino (SM singlet) chiral
  superfields.

\end{itemize}
Excluding the conditions ($b$) and ($d$) for a moment, it was shown in
Table~1 of Ref.~\cite{Dreiner:2003yr} that all 20~$X$-charges of the
MSSM$+\ol{N^i}$ superfields can be expressed in terms of nine real
numbers. Note that for simplicity, we assume three generations of
right-handed neutrinos, unlike in Ref.~\cite{Dreiner:2003yr} where
only two generations were introduced.
\beqn
\begin{array}{lll}
x=0,1,2,3,~~  &   X_{L^1},   & ~~~~~  X_{\overline{N^1}}, \\
y=-1,0,1,~~\, & \Delta_{21}^L \equiv X_{L^2} - X_{L^1}, & ~~~~~
X_{\overline{N^2}},\\
z=0,1,~~~~~~~\: & \Delta_{31}^L \equiv X_{L^3} - X_{L^1}, & ~~~~~
X_{\overline{N^3}},
\end{array}
\label{nonnuconstr} 
\eeqn 
Here $X_F$ denotes the $U(1)_X$-charge of the field $F$.
A few comments are in order:
\begin{itemize}
\item $\Delta_{31}^L$ and $\Delta_{21}^L$ can only take integer values.  
\item $x$ is related to the ratio of the vacuum expectation values 
(VEVs) of the two Higgs doublets, $\tan\beta =\frac{\upsilon_u}{
\upsilon_d}$, by $\eps^x \sim \frac{m_b}{m_t}\tan\beta$.
\item $y$ parameterizes  the phenomenologically viable
  $\eps$-structures for the CKM matrix. Our preferred choice is $y=0$
  as it gives a CKM matrix with $U^{\mathrm{CKM}}_{12} \sim \eps$,
  $U^{\mathrm{CKM}}_{13} \sim \eps^3$, and $U^{\mathrm{CKM}}_{23} \sim
  \eps^2$, see Ref.~\cite{Dreiner:2003yr}.

\item $z$ is related to the ratio $m_e/m_\mu$. It turns out to equal
  $-X_{H^U} - X_{H^D}$ and thus deals with the origin and the
  magnitude of the $\mu$-parameter. For $z=1$, the bilinear Higgs term
  is forbidden before $U(1)_X$-breaking. After $U(1)_X$-breaking it is
  generated via the combination of the FN-mechanism together with the
  GM/KN-mechanism, resulting in a $\mu$-parameter of the order of the
  soft supersymmetry breaking scale $m_{\mathrm{soft}}$.  So the $\mu
  $-problem finds a natural solution, unlike in the case for $z=0$; we
  will hence assume $z=1$ throughout this article.

\item The $X$-charges of the first generation lepton doublet $L^1$ and
  the three right-handed neutrinos are unconstrained at this stage. We
  will explain in a moment why the right-handed neutrinos have to be
  introduced at all.

\item Assuming a string-embedded FN framework, the expansion parameter
  $\eps$ is a derived quantity which depends on $x$ and $z$. For $z=1$
  and $x=0,1,2,3$ we get $\eps$ within the interval (see
  Ref.~\cite{Dreiner:2003yr} and references therein for details)
  \beq\label{epsinterval} 0.186\leq \eps \leq 0.222\,.  
\eeq
\end{itemize} 

Let us now include $(d)$, \ie the constraints arising from
the requirement of a low-energy DGS $P_6$. The necessary and
sufficient conditions on the $X$-charges for obtaining ${P}_6
$~conservation are derived in Ref.~\cite{Dreiner:2006xw}. With $p=\pm
1$ they are 
\beq X_{H^D} - X_{L^1} = -\frac{1}{2}+
\mathrm{integer},~~~~~~~~~ 3X_{Q^1} + X_{L^1} = -\frac{p}{3} +
\mathrm{integer}, 
\eeq 
as well as (see the argument in Item~\ref{itdrei} in
Sect.~\ref{ss31}) the three $X$-charges of the right-handed
neutrinos being half-odd-integer.
Inserting the expression for $X_{Q^1}$ of Table~1 in
Ref.~\cite{Dreiner:2003yr}, we can rewrite this as 
\beq 
\Delta^{H}
\equiv X_{L^1} - X_{H^D}-\frac{1}{2}, ~~~~~~~~~ 3\zeta + p \equiv
\Delta_{21}^L + \Delta_{31}^L - z,\label{p6constraints} 
\eeq 
where $\Delta^H, \zeta \in \mathbb{Z}$. We thus impose proton hexality
by trading the parameters $X_{L^1}$ and $\Delta_{21}^L$ of
Eq.~(\ref{nonnuconstr}) for the \textit{integer} parameters $\Delta^H$
and $3\zeta+p$. The resulting constrained $X$-charges are
shown in Table~\ref{Table2}.
\begin{table}[t!]
\begin{center}
\begin{tabular}{|rcl|}
\hline
  & & \\
$~~~\phantom{\Big|}X_{H^D}$
 &$=$&$\frac{1}{5~(6+x+z )}~\Big(6y + x ~(2x + 11 + z 
- 2\Delta^{\!H}) $\\
       & &$  - z~(\frac{11}{2} + 3 \Delta^{\!H}) - 
  2 ~(6 + 6 \Delta^{\!H} - \Delta^L_{31})-\frac{2}{3}~(6+x+z)(3 \zeta+ p)
\Big)~~~$\\
$~\phantom{\Big|}X_{H^U}$
 &$=$&$-z-\phantom{\Big|}X_{H^D}  $\\
$~\phantom{\Big|}X_{Q^1}$&$=$&$\frac{1}{3}\Big(
      \frac{19}{2} - X_{H^D} + x + 2y + z - 
\Delta^{\!H} - \frac{1}{3}(3 \zeta + p) \Big)$\\
$~\phantom{\Big|}X_{Q^2}$&$=$&$X_{Q^1}-1-y $\\
$~\phantom{\Big|}X_{Q^3}$&$=$&$X_{Q^1}-3-y $\\
$~\phantom{\Big|}X_{\ol{U^1}}$&$=$&$
X_{H^D}-X_{Q^1}+8+z $\\
$~\phantom{\Big|}X_{\ol{U^2}}$&$=$&$
X_{\ol{U^1}}-3+y $\\
$~\phantom{\Big|}X_{\ol{U^3}}$&$=$&$
X_{\ol{U^1}}-5+y $\\
$~\phantom{\Big|}X_{\ol{D^1}}$&$=$&$-X_{
H^D}-X_{Q^1}+4+x $\\
$~\phantom{\Big|}X_{\ol{D^2}}$&$=$&$X_{
\ol{D^1}}-1+y $\\
$~\phantom{\Big|}X_{\ol{D^3}}$&$=$&$
X_{\ol{D^1}}-1+y $\\
$~\phantom{\Big|}X_{L^1}$&$=$&$
X_{H^D}+\Delta^{\!H} + \frac{1}{2}$\\
$~\phantom{\Big|}X_{L^2}$&$=$&$X_{L^1} -
\Delta^L_{31}+ z+ (3\zeta + p) $\\
$~\phantom{\Big|}X_{L^3}$&$=$&$X_{L^1}
+\Delta^L_{31}$\\
$~\phantom{\Big|}X_{\ol{E^1}}$&$=$&$-X_{
H^D}        {+~4} - X_{L^1} + x + z $\\
$~\phantom{\Big|}X_{\ol{E^2}}$&$=$&$X_{\ol{E^1}}
 -2 - 2z + \Delta^L_{31} - (3\zeta  + p)    $ \\ 
$~\phantom{\Big|}X_{\ol{E^3}}$&$=$&$
X_{\ol{E^1}} -4 - \phantom{2}z - \Delta^L_{31}             $ \\ 
$~\phantom{\Big|}X_{\ol{N^1}}$&=&$\frac{1}{2}+\Delta^{\overline{N}}_1$\\
$~\phantom{\Big|}X_{\ol{N^2}}$&=&$\frac{1}{2}+\Delta^{\overline{N}}_2$ \\ 
$~\phantom{\Big|}X_{\ol{N^3}}$&=&$\frac{1}{2}+\Delta^{\overline{N}}_3$ \\ 
  & &\\ \hline
\end{tabular}
\caption{\label{Table2}\small The constrained $X$-charges which lead 
  to an acceptable low-energy phenomenology of quark and charged
  lepton masses and quark mixing. In addition, the GS anomaly
  cancellation conditions have been implemented as well as the
  quadratic anomaly condition. Furthermore, $P_6$ is conserved, \ie
  Eq.~(\ref{p6constraints}) has been imposed. $x$, $y$, $z$ and $p$
  are integers specified in
  Eqs.~(\ref{nonnuconstr},\ref{p6constraints}). $\Delta^H$, $\Delta^L
  _{31}$, and $\zeta$ are integers as well but still
  unconstrained. The $\Delta^{\overline{N}}_i$ of the right-handed
  neutrinos are yet-unspecified integers.}
\end{center}
\end{table}





\cleqn
\section{\label{nusect}Neutrino Constraints 
on the $\boldsymbol{X}$-Charges}
\subsection{\label{ss31}The Origin of $\boldsymbol{P_6}$ 
Neutrino Masses} 

Next we take the remaining constraints $(b)$ into account,
\ie the experimental data from the neutrino sector.  To do
so, let us first consider the possible sources of neutrino masses
in a $P_6$~invariant FN scenario.
\begin{enumerate}
\item Neutrino masses can\textit{not} derive from matter parity
  ($M_p$) violating operators such as $LH^U$ or $LL{\overline E}$,
  as these are forbidden by $P_6$.

\item Therefore, and in the lack of right-handed neutrinos,
  (Majorana) neutrino masses can only originate from the dimension
  five superpotential term $L^iH^UL^jH^U$.  Assuming a minimal number
  of fundamental mass scales, \ie only $m_\mathrm{soft} \approx 0.1
  \,-\, 1 \,\mathrm{TeV}$ and $M_{\mathrm{grav}}=2.4 \cdot 10^{18} \,
  \mathrm{GeV}$, this operator is suppressed by
  $\frac{1}{M_{\mathrm{grav}}}$. This results in the following
  neutrino mass matrix
  \beq\label{kd1}
  \left[\boldsymbol{M^{(\nu)}_{LH^U\!LH^U}}\right]_{ij} ~ \sim ~
  \frac{\langle H^U \rangle^2}{M_{\mathrm{grav}}} \cdot \eps^{X_{L^i}
    + X_{L^j} + 2X_{H^U}}.  
  \eeq 
  Since $X_{L^i} + X_{L^j} + 2X_{H^U} \geq 0$ and $\eps \approx 0.2$,
  the absolute neutrino mass scale cannot exceed $\frac{\langle H^U
  \rangle^2}{M_{\mathrm{grav}}} \approx 1.3\cdot 10^{-5} \mathrm{eV}$
  in this scenario (with $\langle H^U \rangle \sim m_t$).  From the
  observed atmospheric neutrino oscillations, we however know that the
  absolute mass scale must be at least $5\cdot10^{-2}\mathrm{eV}$.
  Thus the neutrino mass matrix cannot (solely) originate from the
  non-renormalizable operator $L^iH^UL^jH^U$. This is not the case if
  we allow for the mass scale which suppresses $L^iH^UL^jH^U$ to be
  lower than $M_{\mathrm{grav}}$, see \eg the model in
  Ref.~\cite{Dreiner:2005rd}. Note that in the case where
  $X_{L^i} + X_{L^j} + 2X_{H^U} < 0$, the operator $L^iH^UL^jH^U$ is
  generated from the K\"ahler potential via the GM/KN-mechanism in
  combination with the FN-mechanism, leading to an even stronger
  suppression by a factor of $\frac{m_{\mathrm{soft}}}{M^2_{
  \mathrm{grav}}}$.

\item \label{itdrei}When enlarging the particle spectrum by three
  generations of right-handed neutrinos $\ol{N^i}$, \ie particles
  which couple trilinearly to $L^iH^U$, a new possibility for the
  neutrino mass term arises.  Since $L^iH^UL^jH^U$ is $P_6$-allowed
  and the term $L^iH^U\ol{N^j}$ by definition as well, but $L^iH^U$ is
  $P_6$-forbidden, the right-handed neutrinos must carry a
  half-odd-integer $X$-charge. Thus the Majorana mass term $\ol{N^i}
  \ol{N^j}$ is necessarily also $P_6$-allowed. In
  Ref.~\cite{Luhn:2007gq}, the possibility of DGSs which allow for
  $L^iH^U\ol{N^j}$ but forbid $\ol{N^i}\ol{N^j}$ and $L^iH^UL^jH^U$
  will be discussed.
\end{enumerate}

Throughout this article, we consider the third possibility above as
the only viable source of neutrino masses in our scenario. The flavon
field $A$ and the right-handed neutrinos $\ol{N^i}$ have a lot in
common. Apart from their $U(1)_X$-charges, both are uncharged. But
there are also certain important differences: 1.)  After $U(1)_X$
breaking $A$ will not carry any $\mathbb{Z}$-charge, whereas the
$\ol{N^i}$ will. 2.) The flavon field $A$ acquires a VEV, whereas the
$\ol{N^i}$ are assumed not to. This is just like the MSSM non-Higgs
scalar fields, which are not supposed to acquire a VEV, in order to
\eg preserve color and/or electromagnetism. Note that $\langle A
\rangle=\eps M_{\mathrm{grav}}$, but $\langle\overline{N^i}\rangle=0$ is 
consistent with the requirement of SUSY being unbroken at $\eps M_
{\mathrm{grav}}$, \ie $\langle D_X\rangle=\langle F_A\rangle=\langle F_{
\overline{N^i}}\rangle=0$.

In the discussion of the constraints on the $X$-charges coming from
the neutrino sector, we have to distinguish between four cases. These
differ in the origin of the superpotential terms $L^iH^U\ol{N^j}$ and
$\ol{N^i} \ol{N^j}$. Depending on the overall $X$-charge, the terms
are either of pure FN origin or effectively generated via the
GM/KN-mechanism in combination with the FN-mechanism. For the Majorana
mass terms, the low-energy effective superpotential terms
are\footnote{We assume that {\it all} entries of the $3 \times 3$ mass
matrices have the same origin: Either they are all generated by
pure FN or all via GM/KN+FN. Allowing otherwise would lead to enormous
suppressions between some of the elements of the mass matrices,
effectively leading to textures, which for simplicity we prefer to
avoid.}
\beqn
X_{\ol{N^i}}+X_{\ol{N^j}}\geq 0:&&
\frac{1}{2} \, M^{\mathrm{(M)}}_{ij} \ol{N^i} \ol{N^j} \;\sim\;
\frac{1}{2}\; M_{\mathrm{grav}} \cdot
\eps^{X_{\ol{N^i}}\,+\,X_{\ol{N^j}}} \cdot \ol{N^i} \ol{N^j}, \label{majFN}\\
X_{\ol{N^i}}+X_{\ol{N^j}}< 0:&& \frac{1}{2} \, M^{\mathrm{(M)}}_{ij}
\ol{N^i} \ol{N^j}\;\sim\; \frac{1}{2}\; m_{\mathrm{soft}} \cdot
\eps^{-X_{\ol{N^i}}\,-\,X_{\ol{N^j}}} \cdot \ol{N^i}
\ol{N^j},\label{majGM} 
\eeqn 
while for the Dirac mass terms we have
\beqn 
X_{L^i}+X_{H^U}+X_{\ol{N^j}}\geq 0:&&
\frac{M^{\mathrm{(D)}}_{ij}}{\langle H^U \rangle} \, L^i H^U
\ol{N^j}\;\sim\; \eps^{X_{L^i}\,+\,X_{H^U}\,+\,X_{\ol{N^j}}}
\cdot L^i H^U \ol{N^j},\label{dirFN}\\
X_{L^i}+X_{H^U}+X_{\ol{N^j}} < 0:&&
\frac{M^{\mathrm{(D)}}_{ij}}{\langle H^U \rangle}\, L^i H^U \ol{N^j}
\;\sim\; \frac{m_{\mathrm{soft}}}{M_{\mathrm{grav}}} \cdot
\eps^{-X_{L^i}\,-\,X_{H^U}\,-\,X_{\ol{N^j}}} \cdot L^i H^U
\ol{N^j}.\label{dirGM}~~~~~~~ 
\eeqn 
The labeling of the four different cases is shown in the following
table.
\begin{center}
\begin{tabular}{|c|c|c|}\hline
 & $\phantom{\Big|}X_{\ol{N^i}} + X_{\ol{N^j}} \geq 0 $ & $X_{\ol{N^i}} 
+ X_{\ol{N^j}}< 0 $\\
 \hline
$\phantom{\Big|}X_{L^i} + X_{H^U} + X_{\ol{N^j}} \geq 0$ & I & II \\ \hline
$\phantom{\Big|}X_{L^i} + X_{H^U} + X_{\ol{N^j}} < 0$ & III & IV\\\hline
\end{tabular}
\end{center}
(This can be compared also to Table~5 of Ref.~\cite{Dreiner:2003yr}:
Case~I contains their~1.+2., Case~II~6., Case~III~3. and
Case~IV~4.+5.)

When determining the masses of the light neutrino degrees of freedom we have
to diagonalize the $6 \times 6$ neutrino mass matrix
\beqn
\begin{pmatrix} \boldsymbol{0} &  \boldsymbol{M^{\mathrm{(D)}}} \\
  {\boldsymbol{M^{\mathrm{(D)}}}}^T & \boldsymbol{M^{\mathrm{(M)}}}
\end{pmatrix}.  \eeqn 
We have approximated the $(1,1)$ entry of the matrix above to be the
$3\times3$ zero matrix, because we already concluded earlier [see
below Eq.~(\ref{kd1})] that $\boldsymbol {M^{(\nu)}_{LH^ULH^U}}$ does
not contribute substantially enough to the absolute neutrino masses.

Under the assumption that the $\eps$-suppression is not able to
compensate the gravitational scale $M_{\mathrm{grav}}$ such that one
arrives at $m_{\mathrm{soft}}$ or $\langle H^U\rangle$ (which would be
$\sim24$ powers of~$\eps$), we see from Eqs.~(\ref{majFN}-\ref{dirGM})
that automatically $\boldsymbol{M^{\mathrm{(D)}}}\ll\boldsymbol{M^{
\mathrm{(M)}}}$ for the Cases I, III and IV. We can thus directly 
apply the see-saw formula to calculate the masses of the three light
neutrinos. In Case II, there are three possibilities
\begin{itemize}
\item[($i$)]  $\boldsymbol{M^{\mathrm{(D)}}} \ll
\boldsymbol{M^{\mathrm{(M)}}}$~~$\longrightarrow$ ~~ standard see-saw,
\item[($ii$)] $\boldsymbol{M^{\mathrm{(D)}}} \approx
\boldsymbol{M^{\mathrm{(M)}}}$,
\item[($iii$)] $\boldsymbol{M^{\mathrm{(D)}}} \gg
\boldsymbol{M^{\mathrm{(M)}}}$ ~~$\longrightarrow$ ~~ pseudo Dirac neutrinos.
\end{itemize}
For Case $(\mathrm{II}.iii)$, the $\eps$-suppression must lower
$\langle H^U \rangle\sim 200 \,\mathrm{GeV}$ down to the neutrino mass
scale, in order to be phenomenologically viable. This corresponds to
about 20 powers of $\eps$ and we do not consider it any further. In
Case~$(\mathrm{II}.ii)$ one would naturally, \ie  without
finetuning among the submatrices $\boldsymbol{M^{\mathrm{(D)}}}$ and
$\boldsymbol{M^{\mathrm{(M)}}}$, expect the neutrino mass matrix to
have six singular values (masses) of {\it the same order}; as for
$(\mathrm{II}.iii)$, extreme $\eps $-suppression is required to obtain
three sub-eV neutrinos. Hence, we also discard Case~($\mathrm{II}.ii
$). For the rest of this article, we refer to Case $(\mathrm{II}.i)$
as Case~II.

Regardless of the Case (I -~IV), in the following the light neutrino
mass matrix is derived from the see-saw mechanism
\cite{{Minkowski:1977sc},{see-saw},{yanagida},Mohapatra:1979ia} and 
is given as (discarding  the contributions from $L^iH^UL^jH^U$)
\beqn
\boldsymbol{M^{(\nu)}} &=&-\boldsymbol{M^{\mathrm{(D)}}} \cdot
\boldsymbol{M^{\mathrm{(M)}}}^{-1} \cdot
{\boldsymbol{M^{\mathrm{(D)}}}}^T. \label{seesaw} 
\eeqn
For later convenience we change the basis of the right-handed
neutrinos so that $\boldsymbol{M^{(\mathrm{M})}}$ is diagonal. Such a basis
transformation is unproblematic after $U(1)_X$ is broken. As discussed
in Ref.~\cite{Dreiner:2006xw}, this basis transformation does not
alter the $\eps$-structure of $\boldsymbol{M^{(\mathrm{D})}}$ in
Eqs.~(\ref{dirFN}) and (\ref{dirGM}). It is now straightforward to
determine $\boldsymbol{M^{(\nu)}}$ for the upper four cases:
\beqn
M^{(\nu,\mathrm{I})}_{ij} &\sim& \frac{\langle
  H^U\rangle^2}{M_{\mathrm{grav}}} \; \eps^{2\Delta^H \,-\,2z\,+\,1 \,+\,
  \Delta^L_{i1} \,+\,\Delta^L_{j1}}, \label{massI}\\  
M^{(\nu,\mathrm{II})}_{ij} &\sim& \frac{\langle
H^U\rangle^2}{m_{\mathrm{soft}}} \; \eps^{2\Delta^H \,-\,2z\,+\,1 \,+\, \Delta^L_{i1} \,+\,\Delta^L_{j1}}  \times \sum_{a=1}^3\eps^{4X_{\ol{N^{a}}}\,}, \label{massII}\\  
M^{(\nu,\mathrm{III})}_{ij} &\sim& \frac{\langle
H^U\rangle^2 \, m_{\mathrm{soft}}^2}{M_{\mathrm{grav}}^3} \; 
\eps^{-\,2\Delta^H  \,+\,2z\,-\,1 \,-\, \Delta^L_{i1}
  \,-\,\Delta^L_{j1}} \times\sum_{a=1}^3\eps^{-\, 4 X_{\ol{N^{a}}}}, \label{massIII}\\  
M^{(\nu,\mathrm{IV})}_{ij} &\sim&  \frac{\langle
H^U\rangle^2 \, m_{\mathrm{soft}}}{M_{\mathrm{grav}}^2} \; 
\eps^{-\,2\Delta^H  \,+\,2z\,-\,1 \,-\, \Delta^L_{i1}
  \,-\,\Delta^L_{j1}}. \label{massIV} 
\eeqn
Here we have made use of Table~\ref{Table2} and the definition  $\Delta^L_{i1}
\equiv X_{L^i} - X_{L^1}$. Note that the dependence on the $X$-charges
of the right-handed neutrinos drops out in Cases~I and~IV, as has been
shown analytically in Ref.~\cite{Dreiner:2003yr}. Thus the masses of
the light neutrinos do not depend on the charges $X_{\ol{N^a}}
$. For Cases~II and~III one might na\"ively expect that although the
{\it overall mass scale} of the light neutrinos depends on the $X_{\ol
{N^a}}$, their {\it mass ratios} $\,\widetilde{m}_3 :\widetilde{m}_2 : 
\widetilde{m}_1 $ do not. The latter however is not true, as is shown 
explicitly for Case~II in Appendix~\ref{detailsect}. Making use of the
orderings\footnote{The ordering of $X_{L^i}$ is necessary for
obtaining a phenomenologically acceptable charged lepton mass matrix
(see the discussion in Ref.~\cite{Dreiner:2006xw}), while we are free
to choose the ordering of $X_{\ol{N^i}}$ without loss of generality.}
$X_{L^3} \leq X_{L^2}\leq X_{L^1}$ and $X_{\ol{N^3}}\leq X_{\ol{N^2}}
\leq X_{\ol{N^1}}$, we obtain
\beq
\widetilde{m}_3 ~:~ \widetilde{m}_2 ~:~ \widetilde{m}_1 ~~\sim~~
1 ~:~ \eps^{2(X_{L^2} - X_{L^3})+4(X_{\ol{N^2}}-X_{\ol{N^3}})} ~:~
      \eps^{2(X_{L^1} - X_{L^3})+4(X_{\ol{N^1}}-X_{\ol{N^3}})}.
\eeq
Assuming $X_{\ol{N^2}} - X_{\ol{N^3}} \geq 1$, the second largest
neutrino mass would be suppressed by a factor of at least $\eps^4$
compared to the heaviest neutrino. Even when including the effects of
unknown $\mathcal{O}(1)$ coefficients, this suppression is too large
to be consistent with the data (see Sect.~\ref {masssection}). For
Case~II, we must therefore constrain the $X $-charges of the
right-handed neutrinos by
\beqn\label{case2hier}
X_{\ol{N^2}}&=&X_{\ol{N^3}},\\ 
\label{case2deg}
X_{\ol{N^1}}&=&X_{\ol{N^2}}~=~X_{\ol{N^3}},
\eeqn
for (normal and inverted) hierarchy and degeneracy, respectively (see
Sect.~\ref{masssection}). 

Similarly for Case~III: Here one obtains the condition 
$X_{\ol{N^1}} = X_{\ol{N^2}}$ for (normal and inverted) hierarchical light 
neutrinos,
and $X_{\ol{N^1}} = X_{\ol{N^2}}= X_{\ol{N^3}}$ for degenerate scenarios.

\subsection{\label{mixingsubsect}Constraints from 
Neutrino Mixing}

The $\eps$-structure of the light neutrino mass matrix is determined
by $\Delta^L_{21}$ and $\Delta^L_{31}$.  We have $M^{(\nu)}_{ij}
\propto \eps^{X_{L^i} + X_{L^j}}$ for Cases I \& II whereas for Cases
III \& IV we find $M^{(\nu)}_{ij} \propto \eps^{-X_{L^i} - X_{L^j}}$.
Both types of matrices are diagonalized by a unitary transformation
$\widetilde U^{(\nu)}_{ij} \sim \eps^{|X_{L^i} - X_{L^j}|}$, so that
\beqn \boldsymbol{{\widetilde U}^{(\nu)}}^\ast \cdot
\boldsymbol{M^{(\nu)}} \cdot
\boldsymbol{\widetilde U^{(\nu)}}^\dagger &=& \begin{pmatrix}
  \widetilde{m}_1 & 0 & 0 \\ 0 & \widetilde{m}_2 &0 \\
  0&0&\widetilde{m}_3
\end{pmatrix},\label{naivetraf}
\eeqn
with 
\beqn
&\mathrm{Case~I:}~~~~~~\;\! & 
\widetilde{m}_1 ~:~ \widetilde{m}_2 ~:~ \widetilde{m}_3 ~\sim~ 1 ~:~ \eps^{2
  \Delta^L_{21}} ~:~ \eps^{2\Delta^L_{31}} \,,
\label{osolemioI} \\
&\mathrm{Case~II:}~~~~~\:\! & 
\widetilde{m}_1 ~:~ \widetilde{m}_2 ~:~ \widetilde{m}_3 ~\sim~ 
1 ~:~ \eps^{2 \Delta^L_{21} +4(X_{\ol{N^2}}-X_{\ol{N^1}})} ~:~ 
\eps^{2\Delta^L_{31}+4(X_{\ol{N^3}}-X_{\ol{N^1}})}  \,,~~~~~~~~
\label{osolemioII} \\
&\mathrm{Case~III:}~~~~ & 
\widetilde{m}_1 ~:~ \widetilde{m}_2 ~:~ \widetilde{m}_3 ~\sim~ 
1 ~:~ \eps^{-2 \Delta^L_{21}-4(X_{\ol{N^2}}-X_{\ol{N^1}})} ~:~
\eps^{-2\Delta^L_{31}-4(X_{\ol{N^3}}-X_{\ol{N^1}})}\,,~~~~~~~~
\label{nessundormaIII} \\
&\mathrm{Case~IV:}~~~~ & 
\widetilde{m}_1 ~:~ \widetilde{m}_2 ~:~ \widetilde{m}_3 ~\sim~ 
1 ~:~ \eps^{-2 \Delta^L_{21}}~:~\eps^{-2\Delta^L_{31}}\,.
\label{nessundormaIV} 
\eeqn
As mentioned above and discussed in greater detail in
Appendix~\ref{detailsect}, the ratios of the light neutrino masses
depend on the $X$-charges of the right-handed neutrinos in Cases~II
and III [see Eqs.~(\ref{rightXII}) and (\ref{rightXIII})].  Recalling
the orderings $X_{L^3} \leq X_{L^2} \leq X_{L^1}$ and $X_{\ol{N^3}}
\leq X_{\ol{N^2}} \leq X_{\ol{N^1}}$ we find
\beq
\mathrm{Cases \;I \,\& \,II}: ~~\widetilde{m}_1 \leq \widetilde{m}_2 
\leq \widetilde{m}_3\,, ~~~~~~~~~~
\mathrm{Cases\;III\, \& \,IV}: ~~\widetilde{m}_1 \geq \widetilde{m}_2 \geq 
\widetilde{m}_3\,,
\eeq
respectively. In order to compare the theoretically derived mixing
matrices $\boldsymbol{\widetilde{U}^{(\nu)}}$ with neutrino
phenomenology, it is convenient to define the matrix $\boldsymbol{U^
{(\nu)}}\equiv \boldsymbol{U^{\mathrm{MNS}}}^{\,\dagger}$, so that
\beqn
\boldsymbol{U^{(\nu)}}^\ast \cdot \boldsymbol{M^{(\nu)}} \cdot
\boldsymbol{U^{(\nu)}}^\dagger &=& \begin{pmatrix} m_1 & 0 & 0
  \\ 0 & m_2 &0 \\ 0&0&m_3 \end{pmatrix}.\label{mnstraf}
\eeqn 
Here $m_1 \leq m_2 \leq m_3$ for normal and $m_3 \leq m_1 \leq m_2$
for inverted ordering of the neutrino masses, see \eg
Ref.~\cite{Giunti:2004vv}. $\boldsymbol{U^{\mathrm{MNS}}}$ is the
Maki-Nakagawa-Sakata matrix~\cite{Maki:1962mu} for mixing in the
lepton sector. Working in a basis with diagonal charged
leptons, {\it cf.} Ref.~\cite{Dreiner:2006xw}, this mixing is solely
due to the neutrino sector. Comparing Eqs.~(\ref{naivetraf},
\ref{mnstraf}), we can easily determine the
relation between $\boldsymbol{U^{(\nu)}}$ and $\boldsymbol{\widetilde
{U}^{(\nu)}}$ and thus the theoretically predicted structure of the 
MNS matrix for the various scenarios:
\begin{itemize}
\item
Considering the Cases I \& II and a normal neutrino mass ordering, we
simply have 
\beqn \boldsymbol{U^{(\nu)}} &=& \boldsymbol{T_{123}} \cdot
\boldsymbol{\widetilde{U}^{(\nu)}}, ~~~~~~\mathrm{with} ~~~~ \boldsymbol{T_{123}}
~\equiv ~ \begin{pmatrix} 1 &0 & 0
  \\ 0& 1&0  \\0 &0 &1    \end{pmatrix},   
\eeqn
\item while 
an inverted mass ordering leads to 
\beqn
\boldsymbol{U^{(\nu)}} &=&
\boldsymbol{T_{231}} \cdot \boldsymbol{\widetilde{U}^{(\nu)}},
~~~~~~\mathrm{with}~~~~\boldsymbol{T_{231}} ~ \equiv ~ \begin{pmatrix} 0 &1 &0
  \\ 0& 0&1  \\1 &0 &0    \end{pmatrix}. 
\eeqn

\item For Cases III \& IV, we similarly find that for a normal neutrino mass ordering 
\beqn
\boldsymbol{U^{(\nu)}} & = & \boldsymbol{T_{321}} \cdot
\boldsymbol{\widetilde{U}^{(\nu)}}, ~~~~~~ \mathrm{with}  ~~~~
\boldsymbol{T_{321}} ~ \equiv ~  \begin{pmatrix} 0 &0 &1 \\ 0& 1&0  \\1 &0 &0
\end{pmatrix} ,
\eeqn
\item and for an inverted mass ordering
\beqn
\boldsymbol{U^{(\nu)}} & = & \boldsymbol{T_{213}} \cdot
\boldsymbol{\widetilde{U}^{(\nu)}}, ~~~~~~ \mathrm{with}  ~~~~
\boldsymbol{T_{213}} ~ 
\equiv ~  \begin{pmatrix} 0 &1 &0 \\ 1& 0&0  \\0 &0 &1    \end{pmatrix}.
\eeqn
\end{itemize}
Since $\boldsymbol{U^{\mathrm{MNS}}}^{\,\dagger}=\boldsymbol{U^{(\nu)}}
= \boldsymbol{T_{...}}\cdot\boldsymbol{\widetilde{U}^{(\nu)}}$, with 
$\widetilde{U}^{(\nu)}_{ij} \sim \eps^{|X_{L^i} -X_{L^j}|}$, we obtain 
severe constraints on the possible values for $\Delta^L_{i1}$ from the 
experimentally allowed $\eps$-structure of the MNS matrix
\cite{Dreiner:2006xw}
\beqn
\boldsymbol{U^{\mathrm{MNS}}} & \sim & \begin{pmatrix}   
\eps^{0,1} & \eps^{0,1} & \eps^{0,1,2,...} \\ 
\eps^{0,1,2} &  \eps^{0,1} & \eps^{0,1} \\ 
\eps^{0,1,2} & \eps^{0,1} & \eps^{0,1} \end{pmatrix}.
\eeqn
Here, multiple possibilities for the exponents of $\eps$ are separated
by commas. Depending on $\boldsymbol{T_{...}}$ we have four different
equations for
\beqn\label{dreizweiO}
\eps^{|X_{L^i} -X_{L^j}|} & \sim & \widetilde{U}^{(\nu)}_{ij} =
\left[ \boldsymbol{T_{...}}^{\dagger} \cdot
\boldsymbol{U^{\mathrm{MNS}}}^{\,\dagger}\right]_{ij} \,.
\eeqn
The resulting $\eps$-structures of $\boldsymbol{\widetilde{U}^{(\nu)}}$ are 
shown in Table~\ref{TableEPS} together with the compatible values for the pairs
$(\Delta^L_{21},\Delta^L_{31})$. 
\begin{table}[t!]
\begin{center}
\begin{tabular}{|c|c|c|}\hline
$\phantom{\Big|}$& Cases I \& II & Cases III \& IV \\ \hline && \\
normal mass ordering & 
$\begin{pmatrix} 
\eps^{0,1} & \eps^{0,1,2} & \eps^{0,1,2} \\ 
\eps^{0,1} & \eps^{0,1} & \eps^{0,1} \\
\eps^{0,1,2,...} & \eps^{0,1} & \eps^{0,1}
\end{pmatrix}$
& 
$\begin{pmatrix} 
\eps^{0,1,2,...} & \eps^{0,1} & \eps^{0,1}\\
\eps^{0,1} & \eps^{0,1} & \eps^{0,1} \\
\eps^{0,1} & \eps^{0,1,2} & \eps^{0,1,2} 
\end{pmatrix}$ \\~&& \\
$(\Delta^L_{21}, \Delta^L_{31})$ & $(0,0),~ (0,-1),~(-1,-2),$ & 
$(0,0),~(0,-1),$ \\
& $(-1,-1)$ & $(-1,-1)$ \\ ~&&\\ \hline ~&& \\ 
inverted mass ordering & 
$\begin{pmatrix} 
\eps^{0,1,2,...} & \eps^{0,1} & \eps^{0,1}\\
\eps^{0,1} & \eps^{0,1,2} & \eps^{0,1,2} \\
\eps^{0,1} & \eps^{0,1} & \eps^{0,1} 
\end{pmatrix}$
&
$\begin{pmatrix} 
\eps^{0,1} & \eps^{0,1} & \eps^{0,1} \\
\eps^{0,1} & \eps^{0,1,2} & \eps^{0,1,2} \\ 
\eps^{0,1,2,...} & \eps^{0,1} & \eps^{0,1}
\end{pmatrix}$ \\ ~&& \\
$(\Delta^L_{21}, \Delta^L_{31})$ & $(0,0),~(0,-1),$ & $(0,0),~(0,-1),$ \\ 
& $(-1,-1)$ & $(-1,-1)$\\
& ~ & ~ \\
\hline
\end{tabular}
\end{center}  \caption{\label{TableEPS}The constraints on the values of
  $\Delta^L_{i1}$ originating from the experimentally observed
  neutrino mixing. The structure of the matrix $\boldsymbol{\widetilde
  {U}^{(\nu)}} =\boldsymbol{T_{...}}^{\dagger} \cdot\boldsymbol{U^{
  \mathrm{MNS}}}^{\,\dagger}$ is shown. As also $\widetilde{U}^{(\nu)
  }_{ij}\sim \eps^{|X_{L^i} -X_{L^j}|}$ must be satisfied, only a few
  pairs of $(\Delta^L_{21}, \Delta^L_{31})$ are possible. Demanding
  $P_6$ invariance, the choice $(-1,-1)$ is excluded, see below
  Eq.~(\ref{dreizweiO}).}
\end{table}
Notice that due to the ordering $X_{L^3} \leq X_{L^2} \leq X_{L^1}$,
we must have $\Delta^L_{i1}\leq 0$ as well as $\Delta^L_{21} \geq
\Delta^L_{31}$.

Having derived the constraints on the parameters $\Delta^L_{i1}$ from
neutrino mixing, we must also satisfy the second condition of
Eq.~(\ref{p6constraints}), which states that $\Delta^L_{21}+\Delta^L_
{31} - z$ must {\it not} be a multiple of three. As mentioned earlier,
we choose to work with $z=1$ in order to have the $\mu$-term generated
by the GM/KN+FN-mechanism. Therefore the choice $(\Delta^L_{21},\Delta
^L_{31}) = (-1,-1)$ is incompatible with the requirement of $P_6$
conservation, and in the remainder of this article we -- of course --
do not consider this $P_6$ violating solution.

We conclude the discussion of the neutrino mixing with some
observations regarding the CHOOZ \cite{Apollonio:1999ae} mixing angle,
$\theta_{13}$. In our notation this angle is parameterized by the
entry $\eps^{0,1,2,...}$ in the mixing matrices $\boldsymbol{
\widetilde{ U}^{(\nu)}}$ of Table~\ref{TableEPS}. As the CHOOZ angle 
is small, one should try to find solutions in terms of $(\Delta^L_{21}
\,,\,\Delta ^L_{31})$ where this entry is $\eps^1$ or $\eps^2$.

Comparing with the four matrices in Table~\ref{TableEPS}, we see that
a normal mass ordering with $(0,-1)$ or $(-1,-2)$ is preferred for
Cases I \& II, while inverted neutrino masses with $(0,-1)$ are
suggested for Cases III \& IV. More precisely, $(0,-1)$
leads to $U^{\mathrm{MNS}}_{13} \sim \eps$ for normal ordered Cases~I
\&~II and inverted ordered Cases~III \&~IV, while $(-1,-2)$
analogously results in $U^{\mathrm{MNS}}_{13} \sim \eps^2$. By
choosing $\Delta^L_{31}$ appropriately, one can understand the
smallness of the CHOOZ angle in terms of the flavor group $U(1)_X$.

There exist of course other possible explanations for the smallness of
$\theta_{13}$. For example, in Ref.~\cite{Shafi:2006nt} this is
achieved by separating the effective neutrino mass matrix as a sum of
two parts; each contains only a $2\times2$ block and is of rank one.
Alternatively, there is a plethora of models adopting non-Abelian
discrete symmetries like \eg $A_4$ \cite{Ma:2001dn,Babu:2002dz,
Zee:2005ut, Altarelli:2005yp}, $\Delta(27)$ \cite{
deMedeirosVarzielas:2006fc, Luhn:2007uq}, $S_3$ \cite{Grimus:2005rf,
Mohapatra:2006pu}, $S_4$ \cite{Brown:1984dk,Hagedorn:2006ug}, $\mathbb
{Z}_7\rtimes\mathbb{Z}_3$ \cite{Luhn:2007sy}, $PSL_2(7)$ 
\cite{Luhn:2007yr} to give rise to the tri-bimaximal mixing 
pattern~\cite{Harrison:2002er}, in which $\theta_{13}$ is exactly zero.

\subsection{\label{masssection}Constraints from Neutrino Masses}

Before discussing the Cases I - IV individually, some general remarks
concerning the magnitude of the three light neutrino masses are in
order. We shall combine the results of the solar \cite{Ahmed:2003kj,
Smy:2003jf}, atmospheric \cite{Ashie:2005ik}, reactor \cite{
Araki:2004mb}, and accelerator \cite{Ahn:2002up} neutrino oscillation
experiments,\footnote{
We disregard the result of the LSND experiment~\cite{
Athanassopoulos:1996jb}, which could not be confirmed by 
MiniBooNE \cite{AguilarArevalo:2007it}.} as well as the upper bound on the
absolute neutrino mass scale originating from the kinematic mass
measurements \cite{Bonn:2001tw}. This leads to three possible
scenarios, see \eg Refs.~\cite{Giunti:2004vv,GonzalezGarcia:2007ib}:
\beqn
m_1 < m_2 \ll m_3 \approx 0.05\,\mathrm{eV}, &&\mathrm{normal~hierarchical},
\notag \\[1mm]
m_3 \ll m_1 < m_2 \approx 0.05 \, \mathrm{eV}, && \mathrm{inverted~
  hierarchical},\notag \\[1mm]
0.05\,\mathrm{eV}  \ll m_1 \approx m_2 \approx m_3 < 2.2\, \mathrm{eV}, 
&& \mathrm{degenerate}.\notag
\eeqn
Assuming a (normal or inverted) hierarchical scenario, the absolute
upper neutrino mass scale $m^\nu_{\mathrm{abs}}\equiv \mathrm{max}
\,(m_1,m_2,m_3)$ is about $0.05~ \mathrm{eV}$, a value which is consistent with
the cosmological upper bound on the sum of the neutrino masses, $\sum_i m_i
\leq 0.7\,\mathrm{eV}$ \cite{Mohapatra:2005wg, Fukugita:2006rm}. 
For an inverted hierarchy, {\it two} neutrinos must have
a mass around this scale, while the third neutrino is much lighter. As
the suppression between the masses of the two heavier neutrinos is
given by [{\it cf.} Eqs.~(\ref{osolemioI}-\ref{nessundormaIV}),
respectively]
\beqn
&\mathrm{Case~I:}~~~~~~\;\! & 
\frac{\widetilde{m}_2}{\widetilde{m}_3} ~\sim~  
\eps^{2  (\Delta^L_{21} - \Delta^L_{31})} \,, \\
&\mathrm{Case~II:}~~~~~\:\! & 
\frac{\widetilde{m}_2}{\widetilde{m}_3} ~\sim~  
\eps^{2  (\Delta^L_{21} - \Delta^L_{31}) + 4(X_{\ol{N^2}} - X_{\ol{N^3}})}
\,,~~~~~~~~\\
&\mathrm{Case~III:}~~~~ & 
\frac{\widetilde{m}_2}{\widetilde{m}_1} ~\sim~  
\eps^{- 2  \Delta^L_{21} + 4(X_{\ol{N^1}} - X_{\ol{N^2}})}
\,,~~~~~~~~\\
&\mathrm{Case~IV:}~~~~ & 
\frac{\widetilde{m}_2}{\widetilde{m}_1} ~\sim~  
\eps^{- 2  \Delta^L_{21} }\,,
\eeqn
the inverted hierarchical scenario is not possible for
all pairs $(\Delta^L_{21},\Delta^L_{31})$:
For Cases~I \&~II we need $(0,0)$ whereas for III \& IV  $(0,0)$ as well
as  $(0,-1)$ are acceptable. 

For the degenerate case, $m^\nu_{\mathrm{abs}}$ can take
values within the range $[0.2\,\mathrm{eV},2.2\,\mathrm{eV}]$, where the lower
end of the interval is estimated such that it satisfies the condition
$0.05\,\mathrm{eV}  \ll m^\nu_{\mathrm{abs}}$. Concerning the cosmological
bound, high values for the neutrino masses are more or less disfavored,
depending on which cosmological observations are included in the derivation of
the bound \cite{Mohapatra:2005wg, Fukugita:2006rm}. We return to this issue in
the discussion of our results. Within our $P_6$ FN-framework, 
the degenerate scenario is only possible if we have $\Delta^L_{21}
=\Delta^L_{31}=0$. This in turn requires a certain amount of finetuning among
the $\mathcal{O}(1)$ coefficients in order to get correct neutrino masses and
mixing. 

We now turn to the discussion of each of the individual Cases I -
IV. In our calculations we take $M_{\mathrm{grav}} = 2.4 \cdot
10^{18}\,\mathrm{GeV}$,
\beq\label{softinterval}
100\,\mathrm{GeV}\leq m_{\mathrm{soft}} \leq 1000 \, \mathrm{GeV},
\eeq 
and $\langle H^U \rangle \sim m_t = 175 \,\mathrm{GeV}.  $\footnote{Of
course, one only knows $\sqrt{\langle H^U \rangle^2+\langle H^D
\rangle^2}$ and not $\langle H^U\rangle$ alone. However, the latter
depends only weakly on $\tan\beta$ (and hence~$x$) in the range
$2\leq \tan\beta\leq50$.} In addition we assume $z=1$, as well
as Eq.~(\ref{epsinterval}).

\begin{itemize}
\item[(I)]
{}From Eq.~(\ref{massI}) and the ordering $\Delta^L_{31} \leq \Delta^
L_{21}\leq \Delta^L_{11} = 0$, we get the absolute neutrino mass scale
as 
\beqn
m^{\nu}_{\mathrm{abs}} &\sim & \frac{m_t^2}{M_{\mathrm{grav}}} \,
\eps^{2\Delta^H + 2\Delta^L_{31} -1}.
\eeqn
Solving for the exponent yields
\beqn
2\Delta^H + 2\Delta^L_{31} -1  &\sim & \frac{1}{\ln{\eps}} \cdot
\ln\left({\frac{m^{\nu}_{\mathrm{abs}}  M_{\mathrm{grav}}}{m_t^2}}\right).
\eeqn 
\begin{itemize}
\item For a normal or inverted hierarchical scenario $m^{\nu}_
{\mathrm{abs}}\approx 0.05\,\mathrm{eV}$. Inserting this and the
limiting values for $\eps$, we arrive at the following allowed range
\beqn
-2\Delta^H - 2\Delta^L_{31} &\in & [3.9 \,,\, 4.5],\label{HIhierpre}
\eeqn
where the lower value of the interval is obtained for small values of
$x$. Since the left-hand side is necessarily an (even) integer, the
hierarchical Case~I slightly prefers small $x$. However, due to
possible unknown $\mathcal{O}(1)$ coefficients we cannot rule out
large~$x$. Furthermore, Eq.~(\ref{HIhierpre}) determines $\Delta ^H$
as
\beqn
\Delta^H & = & -2 - \Delta^L_{31}\,.\label{HIhier}
\eeqn
\item
Considering the degenerate case, which is only possible for $ \Delta^L_{21} =
\Delta^L_{31} =0$, the absolute mass scale
 $m^{\nu}_{\mathrm{abs}}$ should be within the interval
$[0.2\,\mathrm{eV}\,,\,2.2 \,\mathrm{eV}]$. With this we are similarly lead to
\beqn
- 2 \Delta^H & \in & [4.7\,,\,7],
\eeqn
where the lower value corresponds to both small $x$ and small
$m^{\nu}_{\mathrm{abs}}$. Thus we have for the degenerate neutrino scenario
\beqn
\Delta^H & = & -3, \label{HIdeg}
\eeqn
a value which is compatible with all $x=0,1,2,3$. $x=0$ leads to a
neutrino mass scale of $m^{\nu}_{\mathrm{abs}}\approx1.7\,\mathrm{eV}$
and $x=3$ to $m^{\nu}_{\mathrm{abs}} \approx 0.5\,\mathrm{eV}$. Taken
at face value, both are in conflict with the cosmological upper bound
on the sum of the neutrino masses. However, $\mathcal{O}(1)$ coefficients can
alleviate this tension. In the comment column of Table~\ref{48details} we give
the na\"ive sum of the neutrino masses assuming all $\mathcal{O}(1)$
coefficients are exactly one. 
\end{itemize}

All possible sets of parameters $(\Delta^L_{21},\Delta^L_{31},3\zeta+p,
\Delta^H,x)$ are summarized in Table~\ref{parametersI}. The
compatibility with the various neutrino mass scenarios is denoted by
the symbol~$\checkmark$. Note that by virtue of
Eq.~(\ref{p6constraints}), the first three parameters are not
independent of each other. As pointed out earlier, we assume
$z=1$. The allowed values for $y=-1,0,1$ remain unconstrained by the
neutrino sector.  Altogether we can find $4\times4\times3=48$ distinct
sets of $X$-charge assignments (including also less favored
possibilities), which fulfill the constraints of
Tables~\ref{Table2}+\ref{parametersI}.  They are given in
Appendix~\ref{tablesect}, Table~\ref{48sets}.
\begin{table}
\begin{center}
\begin{tabular}{|c|c|c|c|c|c|c|c|}\hline
$\phantom{\Big|}\Delta^L_{21}$ & $\Delta^L_{31}$ & $3\zeta + p$ & $\Delta^H$
& $x$ &  normal hier. & inverted hier. & degenerate \\ \hline
$\phantom{\Big|}0$ & $0$ & $-1$ & $-3$ & $0,1,2,3$ & &  & $\checkmark$
\\ \hline 
$\phantom{\Big|}0$ & $0$ & $-1$ & $-2$ & $0,1,(2,3)$ &  $\checkmark$ & 
$\checkmark$ &
\\ \hline 
$\phantom{\Big|}0$ & $-1$ & $-2$ & $-1$ & $0,1,(2,3)$ &  $\checkmark$ & 
& \\ \hline  
$\phantom{\Big|}-1$ & $-2$ & $-4$ & $0$ & $0,1,(2,3)$ &  $\checkmark$ & & \\
\hline
\end{tabular}
\end{center}\caption{\label{parametersI}The sets of parameters which are
  compatible with neutrino phenomenology in  Case I, where the terms
  $L^iH^U\ol{N^j}$ and $\ol{N^i}\ol{N^j}$ have pure FN origin. 
  We assume $z=1$. The hierarchical scenarios slightly prefer small $x$ and
  disfavor large (denoted by the parentheses). 
  The parameter $y=-1,0,1$ remains   unconstrained.}
\end{table}

For  Case~I, the $X$-charges of the right-handed neutrinos are not directly
constrained by neutrino phenomenology. Recall however that this case requires
by definition  $X_{L^i} + X_{H^U} + X_{\ol{N^j}} \geq 0$ and $X_{\ol{N^i}} +
  X_{\ol{N^j}} \geq 0$ for all $i,j =1,2,3$. With Table~\ref{Table2} 
  and $z=1$  this translates into 
\beqn
\Delta^{\ol{N}}_i & \geq & -\Delta^L_{31} - \Delta^H, \label{NlowI}
\eeqn
leading to $\Delta^{\ol{N}}_i \geq 3$ in the degenerate case and $
\Delta^{\ol{N}}_i \geq 2$ for hierarchical scenarios.

On the other hand, there exists also an upper bound on
$\Delta^{\ol{N}}_i$.  Qualitatively, very high $X$-charge for the
right-handed neutrinos would suppress the Majorana mass matrix in
Eq.~(\ref{majFN}) so that its mass scale becomes comparable to or even
smaller than the Dirac masses of Eq.~(\ref{dirFN}). Thus the see-saw
formula would no longer apply. Requiring that $M^{\mathrm{(D)}}_{33}
\ll M^{\mathrm {(M)}}_{11}$ yields the condition
\beqn\label{NupI}
2\Delta^{\ol{N}}_1 -\Delta^{\ol{N}}_3 &<&
\frac{1}{\ln{\eps}} \cdot \ln\left({\frac{\langle H^U \rangle}
{M_{\mathrm{grav}}}}\right)  
\, - \, z + \Delta^L_{31} + \Delta^H.
\eeqn
Depending on $\eps$, the first term on the right-hand side is
numerically between $22.1$ and $24.7$. With the latter, \ie for the
case where $\eps=0.222$, we arrive at the upper bounds of $2\Delta^{
\ol{N}}_1 -\Delta^{\ol{N}}_3 \leq 20$ for the degenerate and $2\Delta
^{\ol{N}}_1 -\Delta^{\ol{N}}_3 \leq 21$ for the hierarchical case,
respectively. In Sect.~\ref{anomaliesetc}, we will constrain the 
$\Delta^{\ol{N}}_i$ by requiring the absence of $X$-charged hidden
sector superfields.

  It is worth noting that thermal leptogenesis requires the
  lightest right-handed neutrino to be not too light: $M^{(\mathrm{M})}_{11}
  \gtrsim 4 \times 10^8$~GeV if the spectrum is hierarchical (no close states)
  but otherwise with rather conservative assumptions~\cite{Buchmuller:2002jk}.
  Even though the considerations here do not determine the $X$-charges of
  $\overline{N^i}$, and hence their masses, we do obtain quite restrictive
  constraints once we require that all
  anomalies are canceled without introducing additional (hidden)
  fields charged only under $U(1)_X$ but not the standard model. See
  Appendix~\ref{tablesect} for more details.

\item[(II)] Proceeding with Case II, we obtain from Eq.~(\ref{massII}) and the
orderings  $\Delta^L_{31} \leq \Delta^L_{21}
\leq \Delta^L_{11} = 0$ and
$X_{\ol{N^3}} \leq X_{\ol{N^2}} \leq X_{\ol{N^1}}$ that 
\beqn
m^{\nu}_{\mathrm{abs}} &\sim & \frac{m_t^2}{m_{\mathrm{soft}}} \, 
 \eps^{2\Delta^H + 2\Delta^L_{31} -1 + 4 X_{\ol{N^3}}}.\label{IIfirst}
\eeqn
The hierarchical scenarios require 
\beqn
2 \Delta^H + 2 \Delta^L_{31} + 4 X_{\ol{N^3}} & \in & [17.1\,,\,20.6],
\label{examplecond} 
\eeqn
the left boundary of the interval corresponds to small $\eps$ [see
Eq.~(\ref{epsinterval})] and {\it large} $m_{\mathrm{soft}}$ [see
Eq.~(\ref{softinterval})]. Thus $\Delta^H$ is given by
\beqn
\Delta^H & = & -\Delta^L_{31} - 2 X_{\ol{N^3}} +  \left\{ \begin{array}{cl} 9,
    & ~~~~x=0,1,(2)\,,
    \\ 10, &~~~~x=2,3\,. \end{array} \right.\label{absIIh}
\eeqn
Here and in the following, values in parentheses are acceptable only
if we rely on suitable $\mathcal{O}(1)$ coefficients to satisfy
phenomenological conditions similar to Eq.~(\ref{examplecond}) with
the above specified parameter ranges. For instance, without any
$\mathcal{O}(1)$ coefficients in Eq.~(\ref{IIfirst}), the value
$x=(2)$ leads to $m_{\mathrm{soft}} = 1990\,\mathrm{GeV}$ which is
outside of the initially assumed range for the soft supersymmetry
breaking scale.

The three possible values of $x$ in the first line of
Eq.~(\ref{absIIh}) yield $m_{\mathrm{soft}}\approx 230\,\mathrm{GeV}$,
$m_{\mathrm{soft}}\approx 680\,\mathrm{GeV}$, and $m_{\mathrm{soft}}
\approx 1990\,\mathrm{GeV}$, respectively. For the second line, we find
analogously $ 90 \,\mathrm{GeV}$ and $ 230\,\mathrm{GeV}$, for
$x=2,3$. As pointed out above, these ``predictions'' of the soft
supersymmetry breaking scale do not take into account the variation
due to the unknown $\mathcal{O}(1)$ coefficients in any FN
model. Allowing for such a factor to be anything within the interval
$[\frac{1}{\sqrt{10}}\,,\,\sqrt{10}]$, there is actually no hard
constraint on $m_{\mathrm{soft}}$, except for the case with $x=2$
which prefers large $m_{\mathrm{soft}}$ in the first line and low
$m_{\mathrm{soft}}$ in the second.

For degenerate neutrinos, the possible variation of the absolute mass
scale within the interval $[0.2\,\mathrm{eV}\,,\,2.2\,\mathrm{eV}]$
leads to a further widening of the allowed range for~$\Delta^H$, in
addition to flexibility in $\eps$ and $m_{\mathrm{soft}}$. Since $
\Delta^L_{21} =\Delta^L_{31} = 0$, we have
\beqn
2 \Delta^H  + 4 X_{\ol{N^3}} & \in & [14.9\,,\,19.6],
\eeqn 
which results in the possible values
\beqn
\Delta^H & = &  - 2 X_{\ol{N^3}} +  \left\{ \begin{array}{ll} 8,
    & ~~~~x=0,1,2,(3),
    \\ 9, &~~~~x=1,2,3. \end{array} \right.\label{absIId}
\eeqn
Again, there is no significant constraint on $m_{\mathrm{soft}}$.
However, the first line of Eq.~(\ref{absIId}) with $x=2,(3)$ prefers a
large soft breaking scale while the second line with $x=1$ suggests
low $m_{\mathrm{soft}}$. Due to the constraints on the $U(1)_X
$-charges given in Table~\ref{Table2}, we can define an integer $n$ as
\beqn
n &\equiv & - X_{\ol{N^3}} -\frac{1}{2}. \label{defofn}
\eeqn
Since $X_{\ol{N^i}}+ X_{\ol{N^j}} < 0 $, the $X$-charges of the
right-handed neutrinos must be negative, hence $n \geq 0$. Another
condition is that
\beq
X_{L^i} + X_{H^U} + X_{\ol{N^j}} = \Delta^L_{i1} +
\Delta^H -\frac{1}{2} + X_{\ol{N^j}}\geq\Delta^L_{31} + \Delta^H -
\frac{1}{2} + X_{\ol{N^3}} \geq 0\,. 
\eeq
Inserting respectively Eqs.~(\ref{absIIh},\ref{absIId}) shows that
this is automatically satisfied. However, there is yet another
relation to be met. Recall that for the see-saw mechanism we require
$\boldsymbol{M^{\mathrm{(D) }}}\ll\boldsymbol{M^{\mathrm{(M)}}}$.
This provides us with a lower bound on $X_{\ol{N^i}}$, as can be seen
in the following.  From Eqs.~(\ref{majGM}) and (\ref{dirFN}), the
lightest right-handed neutrino has a Majorana mass of the order
$m_{\mathrm{soft}}\,\eps^{-\,2X_{\ol{N^3}}} $ and the heaviest Dirac
mass is of order $\langle H^U\rangle\,\eps^{X_{L^3} + X_{H^U} +
X_{\ol{N^3}}}$. Therefore we require
\beqn
\frac{m_{\mathrm{soft}}}{\langle H^U \rangle}& \gg & \eps^{X_{L^3} + X_{H^U} + 3
  X_{\ol{N^3}}}\,.
\eeqn
As a conservative estimate, we take $m_{\mathrm{soft}}=1000\,\mathrm
{GeV}$, yielding $\frac{m_{\mathrm{soft}}}{\langle H^U\rangle}\approx 
\eps^{-1}$ for the left-hand side. Therefore
\beqn
\Delta^L_{31} + \Delta^H -\frac{1}{2} + 3  X_{\ol{N^3}}  = X_{L^3} +
 X_{H^U} + 3  X_{\ol{N^3}} &
> &-1. \label{RHX}
\eeqn 
For the hierarchical cases, we insert Eq.~(\ref{absIIh}) into
Eq.~(\ref{RHX}).  Expressing $X_{\ol{N^3}}$ in terms of $n \geq 0$ we
arrive at the conditions
\beqn
 0 \leq n \leq \left\{ \begin{array}{ll} 8, ~~&~~ x=0,1,(2), \\ 9, ~~& ~~
     x=2,3, \end{array} \right. 
\eeqn
where the two lines correspond to the two possibilities for $\Delta^H$ in
Eq.~(\ref{absIIh}). 

For the degenerate case, where $\Delta^L_{31}=0$, we similarly obtain with
Eq.~(\ref{absIId}) 
\beqn
 0 \leq n \leq \left\{ \begin{array}{ll} 7, ~~&~~ x=0,1,2,(3), \\ 8, ~~& ~~
     x=1,2,3. \end{array} \right. 
\eeqn
In Table~\ref{parametersII}, we give all sets of parameters $(\Delta^L
_{21},\Delta^L_{31},3\zeta + p , \Delta^H,x,n)$, which comply with the
phenomenology of neutrino masses and mixings for Case II. We assume
$z=1$, and the parameter $y=-1,0,1$ remains unaffected by the neutrino
sector. Compared to the analogous table for Case~I, we have added the
parameter $n \in \mathbb{N}$, which is defined by the $X$-charge of
the right-handed neutrino $\ol{N^3}$ [{\it cf.} Eq.~(\ref{defofn})]
and determines the parameter $\Delta^H$. Limiting ourselves to Case II
restricts the allowed values for~$n$.
\begin{table}
\begin{center}
\begin{tabular}{|c|c|c|c|c|c|c|c|c|}\hline
$\phantom{\Big|}\!\!\!\Delta^L_{21}$ & $\!\Delta^L_{31}\!$ & $\!3\zeta + p\!$ 
& $\Delta^H$ & $x$ & $n$ 
& $\!$normal$\!$ & $\!$inverted$\!$ &$\!$degenerate$\!$ \\ \hline
$\phantom{\Big|}0$ & $0$ & $-1$ & 
$\phantom{\Bigg|} \!\!\! 2n + \left\{ \begin{array}{c}  9\\10 \end{array}\right.\!\!\!\!$ & 
$\phantom{\Bigg|}\!\!\!\!\begin{array}{l} 0,1,2,(3)\\~~1,2,3\end{array} \!\!\!$ &
$\phantom{\Bigg|} \!\!\!\! \begin{array}{l}  0 \leq n \leq 7 \\ 0 \leq n \leq 8
                   \end{array}\!\!\!$ &
    &  &  $\checkmark$ \\ \hline
$\phantom{\Big|}0$ & $0$ & $-1$ & 
$\phantom{\Bigg|} \!\!\! 2n + \left\{ \begin{array}{c}  10\\11 \end{array}\right.\!\!\!\!$ & 
$\phantom{\Bigg|} \!\!\!\! \begin{array}{l}  0,1,(2)\\~~2,3 \end{array} \!\!\!$ &
$\phantom{\Bigg|} \!\!\!\! \begin{array}{l}  0 \leq n \leq 8 \\ 0 \leq n \leq 9
                   \end{array}\!\!\!$ &
  $\checkmark$  &  $\checkmark$  &   \\ \hline
$\phantom{\Big|}0$ & $-1$ & $-2$ & 
$\phantom{\Bigg|} \!\!\! 2n + \left\{ \begin{array}{c}  11\\12 \end{array}\right.\!\!\!\!$ & 
$\phantom{\Bigg|} \!\!\!\! \begin{array}{l}  0,1,(2)\\~~2,3 \end{array} \!\!\!$ &
$\phantom{\Bigg|} \!\!\!\! \begin{array}{l}  0 \leq n \leq 8 \\ 0 \leq n \leq 9
                   \end{array}\!\!\!$ &
  $\checkmark$  &    &   \\ \hline
$\phantom{\Big|}-1$ & $-2$ & $-4$ & 
$\phantom{\Bigg|} \!\!\! 2n + \left\{ \begin{array}{c}  12\\13 \end{array}\right.\!\!\!\!$ & 
$\phantom{\Bigg|} \!\!\!\! \begin{array}{l}  0,1,(2)\\~~2,3 \end{array} \!\!\!$ &
$\phantom{\Bigg|} \!\!\!\! \begin{array}{l}  0 \leq n \leq 8 \\ 0 \leq n \leq 9
                   \end{array}\!\!\!$ &
  $\checkmark$  &    & \\\hline
\end{tabular}
\end{center}\caption{\label{parametersII}The sets of parameters which
 are compatible with neutrino phenomenology in Case II where the term
 $L^iH^U\ol{N^j}$ has pure FN origin while $\ol{N^i}\ol{N^j}$ is
 generated via GM/KN.  We assume $z=1$. The parameter $y=-1,0,1$
 remains unconstrained, $n$ can take only positive integer values
 which are restricted as shown in the table.}
\end{table}
Altogether we can thus find $[(4\times8+3\times9)+(3\times9+2\times10)
+(3\times9+2\times10)+(3\times9+2\times10)]\times3=600$ sets of $X
$-charge assignments, including also less favored possibilities.  Some
of these charge assignments, however, are identical due to the first
two rows of Table~\ref{parametersII}. There are 504 distinct
sets of $X$-charges.  A selected subset resulting from
Table~\ref{parametersII} is given in Appendix~\ref{tablesect},
Table~\ref{24sets}. For the relevant criteria see the next Section.

  It is worth noting that there is a constraint from neutrino
  oscillation due to the presence of right-handed states.  The most
  stringent limit comes from appearance experiment searches:
  $\nu_\mu \rightarrow \nu_{\ell}$, for $\ell = e, \tau$.  
  Because we required the right-handed neutrino masses to be much higher than
  the light neutrino masses, Eq.~(3.48), the oscillation probability is
  averaged out, and hence we obtain an upper limit on the mixing angle. The
  effective  ``$\sin^2 2\theta$'' must be less than about $3\times 10^{-4}$ 
  \cite{Eskut:2000de,Astier:2001yj,Armbruster:2002mp,Astier:2003gs,AguilarArevalo:2007it}. 
  In  terms of the mixing matrices, the limit is therefore $|U_{\mu i} U_{\ell    i}^*| \approx \frac{M^{(\mathrm{D})}_{\mu i} M^{(\mathrm{D})}_{\ell
  i}}{{M^{(\mathrm{M})}_{ii}}^2} \lesssim \epsilon^2$, where we assumed
  $m_{\mathrm{soft}} \sim \langle H^U \rangle$.  Therefore, we obtain $2
  \Delta_H + \Delta_{21}^L +  \Delta^L_{\ell i}-6n~>~5$.  This restricts the
  allowed ranges of $n$ in Table 4 slightly more: All upper limits on $n$ are
  reduced by 1 to 6, 7, 7, 8, 7, 8, 7, 8, respectively.

%
%
%
\item[(III)]
For Case~III the scale of the Dirac mass matrix
$M^{(\mathrm{D})}_{ij}$ in Eq.~(\ref{dirGM}) is given by the $(1,1)$
entry. Since  $X_{L^1} + X_{H^U} + X_{\ol{N^1}} < 0$, this mass
scale has an upper bound
\beqn
M^{\mathrm{(D)}}_{11} & < & \frac{\langle H^U
  \rangle \,m_{\mathrm{soft}}  }{M_{\mathrm{grav}}} \,. 
\eeqn
Calculating the light neutrino mass matrix by the see-saw formula,
Eq.~(\ref{seesaw}), can only generate an absolute neutrino mass scale
$m^{\nu}_{\mathrm{abs}}$ which is smaller than $M^{(\mathrm{D})}_{11}
$. Furthermore,
\beqn
 0.05 ~\mathrm{eV}  & \leq  & m^{\nu}_{\mathrm{abs}}~~ <~~
 M^{\mathrm{(D)}}_{11}  ~~ < ~~  \frac{\langle H^U
  \rangle \, m_{\mathrm{soft}} }{M_{\mathrm{grav}}}\,,
\eeqn
and thus the soft scale has to be extraordinarily large, at least $500
\,\mathrm{TeV}$. This renders Case~III highly unattractive. We will
therefore not elaborate on the possibility of the Dirac mass matrix
being generated by GM/KN+FN any further.

\item[(IV)] As for Case~III.

\end{itemize}





\cleqn

\section{\label{anomaliesetc}An $\boldsymbol{X}$-charged Hidden Sector?}

The GS cancellation of chiral anomalies often requires the introduction
of further $X$-charged matter fields, which are singlets under the
Standard Model gauge group, \ie hidden sector superfields, for
examples see Refs.~\cite{Dreiner:2003yr,Dreiner:2006xw,
Chamseddine:1995gb}.\footnote{However, in Ref.~\cite{Dreiner:2003yr}, with
three instead of two generations of right-handed neutrinos and $k_C = 3$ the
GS anomaly cancellation conditions could also have been satisfied without
exotic matter.}  
But as we now explain, in our $P_6$ conserving FN
study, it is possible to have GS anomaly cancellation without exotic,
hidden sector, matter. In such a case, anomaly considerations open up
a window on the underlying string theory. It should be stressed that
the condition of no further $X$-charged matter is an option which does
not affect any of the previous considerations.

Two of the GS conditions are given as\footnote{We differ from
Ref.~\cite{Dreiner:2003yr} by a factor of 3 in the denominator of the
third ratio.} \cite{Dreiner:2003yr}
\beq
\frac{\mathcal{A}_{CCX}}{k_C}=\frac{\mathcal{A}_{GGX}}{24} 
= \frac{\mathcal{A}_{XXX}}{k_X},\label{GScond}
\eeq 
where the positive real parameters $k_{...}$ are the affine or
Ka\v{c}-Moody levels, which take integer values for non-Abelian gauge
groups.  $\mathcal{A}_{...}$ denote the anomaly coefficients, with $G$
standing for ``gravity'', $C$ for $SU(3)_C$, and $X$ for $U(1)_X$. The
$k_{...}$ are related to the corresponding gauge coupling constants
at the unification scale
\beq
g_C^2k_C=g_X^2k_X=2g_{\mathrm{string}}^2.\label{gstring}
\eeq
These $2+2$ equations give
\beq\label{ooooo}
\begin{array}{rclrcl}
g_{\mathrm{string}}&=&g_C\sqrt{\displaystyle{\frac{12\cdot A_{CCX}}{A_{GGX}}}},&
\qquad g_X&=&g_C
\sqrt{\displaystyle{\frac{A_{CCX}}{A_{XXX}}}}\;,\\[5mm]
\qquad k_X&=&\displaystyle{\frac{24\cdot A_{XXX}}{A_{GGX}}},
&k_C&=&\displaystyle{\frac{24\cdot A_{CCX}}{A_{GGX}}}\;.
\end{array}
\eeq
Assuming, as in deriving Table~\ref{Table2}, that all non-MSSM
superfields are color singlets, we have
\beqn
\mathcal{A}_{CCX}&=&\mbox{$\frac{1}{2}$}\sum_{i=1}^3
\left(2X_{Q^i}+X_{\ol{U^i}}+X_{\ol{D^i}}\right),\\
\mathcal{A}_{GGX}&=&\sum_{i=1}^3 
\left(6X_{Q^i}+3X_{\ol{U^i}}+
3X_{\ol{D^i}}+2X_{L^i}+X_{\ol{E^i}}+X_{\ol{N^i}}\right)\\
&~& +~ 2\left(X_{H^D}+X_{H^U}\right)~+~X_A+\mathcal{A}_{GGX}^{\mathrm{hidden}},
\nonumber\\
\mathcal{A}_{XXX}&=& \sum_{i=1}^3 
\left(6{X_{Q^i}}^3+3{X_{\ol{U^i}}}^3+
3{X_{\ol{D^i}}}^3+2{X_{L^i}}^3+{X_{\ol{E^i}}}^3
+{X_{\ol{N^i}}}^3\right)\\
&~& +~2\left({X_{H^D}}^3+{X_{H^U}}^3\right)~+~
{X_A}^3+\mathcal{A}_{XXX}^{\mathrm{hidden}}.\nonumber
\eeqn
Here and in Eq.~(\ref{gstring}), we have used the standard
GUT-normalization of non-Abelian groups with generators $t_a$ such
that $\mathrm{tr}[t_at_b]=\frac{1}{2}{\delta_{ab}}$. With Table~1, we
get \eg
\beqn
\mathcal{A}_{CCX}&=&\mbox{$\frac{3}{2}$}(6+x+z),\\
\mathcal{A}_{GGX}&=&62 + 12 x + 
8 z + {\Delta_1^{\ol{N}}+ \Delta_2^{\ol{N}}+
\Delta_3^{\ol{N}}}  
+  \Delta_{21}^L +   \Delta_{31}^L + 
3 \Delta^H+\mathcal{A}_{GGX}^{\mathrm{hidden}}.~~
\eeqn
So despite the 17 MSSM $X$-charges being known, \textit{cf.}
Tables~\ref{parametersI} and~\ref{parametersII}, we cannot give
numerical values for $\{g_{\mathrm{string}},~g_X,~k_X,~k_C\}$, since
the $\Delta_i^{\ol{N}}$, $\mathcal{A}_{GGX}^{\mathrm{hidden}}$ and
$\mathcal{A}_{XXX}^{\mathrm{hidden}}$ are still unknown. \textit{But}
now let us suppose that the left-chiral MSSM superfields, as well as
the $\ol{N^i}$ and the flavon $A$ are the only $X$-charged
superfields. Hence $\mathcal{A}_{GGX}^{\mathrm{hidden}}$ and $\mathcal
{A}_{XXX}^{\mathrm{hidden}}$ vanish.\footnote{$\mathcal{A}_{GGX}^{
\mathrm{hidden}}$ and $\mathcal{A}_{XXX}^{\mathrm{hidden}}$ also 
vanish if the additional exotic particles are vector-like.}  We can
then scan all 48+504 $X$-charge assignments, defined by the parameters
$\{x,z,\Delta_{21}^L,\Delta_{31}^L,\Delta^H\} $, for solutions to the
fourth equality of Eq.~(\ref{ooooo}) with the requirement of {\it $k_C$ being
  an integer}: 
\beq
k_C= \frac{36(6+x+z)}{62 + 12 x + 
  8 z + {\Delta_1^{\ol{N}}+ \Delta_2^{\ol{N}}+
\Delta_3^{\ol{N}}}+  \Delta_{21}^L +   \Delta_{31}^L + 3 \Delta^H} \,. 
\label{noexotics}
\eeq
As pointed out above, the integers $\Delta_ i^{\ol{N}}$ are already
constrained. Besides the required ordering $ \Delta_3^{\ol{N}}\leq 
\Delta_2^{\ol{N}}\leq\Delta_1^{\ol{N}}$ we have
\begin{itemize}
\item For Case~I, see below Eqs.~(\ref{NlowI}) and (\ref{NupI}),
\beqn
&&\mathrm{hierarchical:}~~~~~~~~2 \leq \Delta_3^{\ol{N}} \,,~~~~~ 
2\Delta_1^{\ol{N}} - \Delta_3^{\ol{N}} \leq 21\,, \label{NhierI}\\
&&\mathrm{degenerate:}~~~~~~~\!\!\:\,~~3 \leq \Delta_3^{\ol{N}} \,,~~~~~ 
2\Delta_1^{\ol{N}} - \Delta_3^{\ol{N}} \leq 20\,,~~~~~~~ \label{NdegI}
\eeqn
\item and for Case~II, see Eqs.~(\ref{case2hier},\ref{case2deg},\ref{defofn}),
with $n$ given in Table~\ref{parametersII},
\beqn
&&\mathrm{hierarchical:}~~~~~~~~ -n-1 \,=\, \Delta_3^{\ol{N}}
=\Delta_2^{\ol{N}}\leq \Delta_1^{\ol{N}} < 0\,, \label{NhierII}\\
&&\mathrm{degenerate:}~~~~~~~\!\!\:\,~~-n-1 \,=\, \Delta_3^{\ol{N}}
=\Delta_2^{\ol{N}} = \Delta_1^{\ol{N}}\,.\label{NdegII}
\eeqn
\end{itemize} 
We then find that the 48 sets of Case~I are all in accord with
$k_C=3$. The required values for $\sum_i \Delta_i^{\ol{N}}$ are given
in Table~\ref{48details}. The conditions on $\Delta_i^{\ol{N}}$
however do {\it not} determine the $X$-charges of the right-handed
neutrinos uniquely; see Appendix~\ref{tablesect} for a complete list
of the remaining possibilities in each case.  On the other hand, there
exist six cases ($\#~25,26,27,37,38,39$) which are also compatible
with $k_C=2$. In these models, the constraints on $\Delta_i^{\ol{N}}$
fix their individual values uniquely, {\it cf.}
Table~\ref{48details}.

Turning to Case~II, the $X$-charges of the right-handed neutrinos 
have to satisfy stronger constraints due to
Eqs.~(\ref{NhierII},\ref{NdegII}). Demanding Eq.~(\ref{noexotics}), 
only 24 of the 504 
models in Table~\ref{parametersII} survive; they are displayed in
Table~\ref{24sets}. In all 24 cases we have $k_C=2$, and
$\Delta_i^{\ol{N}}$ is fixed uniquely as given in Table~\ref{24details}. 

A brief comment about the number of possible models before and after
imposing Eq.~(\ref{noexotics}) is in order. Excluding the right-handed
neutrinos, we start with 48 distinct sets of $X$-charge assignments in
Case~I and 504 in Case~II. This huge difference is due to the fact
that in Case~II the dependence of the effective neutrino mass matrix
$\bsym{M^{(\nu)}}$ on the right-handed neutrinos $\overline{N^i}$, see
Eq.~(\ref{massII}), allows for a variation of $\Delta^H$ parameterized
by $n$. In Case~I, such a dependence and thus a similar parameter is
absent. Taking the right-handed neutrinos into account, the dependence
of $\bsym{M^{(\nu)}}$ on $\overline{N^i}$ strongly limits the possible
$X$-charges for $\overline{N^i}$ in Case~II [{\it cf.}
Eqs.~(\ref{NhierII},\ref{NdegII})], whereas for Case~I, $X_{\overline
{N^i}}$ can be chosen from an interval [{\it cf.} 
Eqs.~(\ref{NhierI},\ref{NdegI})]. When it comes to finding solutions
to Eq.~(\ref{noexotics}), this freedom of assigning $X_{\overline{N^i}
}$ in Case~I allows each of the 48 sets of $X$-charges to be
consistent without an $X$-charged hidden sector. In Case~II, the
situation is much more constrained, reducing 504 models to only 24
viable ones.

Having determined the Ka\v{c}-Moody levels $k_C$ which are consistent
with the assumption of no exotic $X$-charged matter, we can calculate
the string coupling constant $g_{\mathrm{string}}$ from
Eq.~(\ref{gstring}). Inserting $g_C[M_{\mathrm{string}}]\approx g_C[
M_{\mathrm{GUT}}]= 0.72$ we get
\beqn
g_\mathrm{string} &\sim &0.88\,,~~~~~~\mathrm{for}~~k_C=3\,, \\
g_\mathrm{string} &\sim &0.72\,,~~~~~~\mathrm{for}~~k_C=2\,. 
\eeqn
From $k_C$ we can obtain the other Ka\v{c}-Moody levels of $G_{
\mathrm{SM}}$ from the gauge coupling unification
relation\footnote{A non-standard gauge coupling unification with $k_C
=k_W =\frac{3}{4} k_Y$ was put forward in Refs.~\cite{Barger:2005gn,
Barger:2005qy} and has been recently applied to FN models in
Ref.~\cite{Gogoladze:2007ck}.}
\beq
k_C ~=~k_W~=~\frac{3}{5} \, k_Y \,,
\eeq
which adopts the $Y$-normalization with $Y_L=1/2$, and has already
been implemented when deriving Table~\ref{Table2}, {\it cf.}
Ref.~\cite{Dreiner:2003yr}.  Thus, the models of Case~I with $k_C=3$
have $k_W=3$ and $k_Y=5$, while those with $k_C=2$ (\ie six models of
Case~I and all models of Case~II) demand $k_W=2$ and $k_Y=10/3$. 

The question arises whether Ka\v{c}-Moody levels $k_C$ and
$k_W$ higher than 1 can be obtained from string model
building. Actually, such models have been considered, \eg
\cite{Lewellen:1989qe,Aldazabal:1994zk,Dienes:1995sq}, but a
systematic investigation of this issue is lacking. Nevertheless, there
are indications that higher Ka\v{c}-Moody levels might occur rather
generically, see \eg Ref.~\cite{LopesCardoso:1994ik}.  Also, from the
phenomenological point of view, models with higher levels have already
been discussed, \eg in Ref.~\cite{Babu:2002tx}.  This is important
regarding the possible representations for the Higgs fields in the
theory \cite{Dreiner:1988zf,Dreiner:1988ax, Antoniadis:1989zy}.

In addition to the Ka\v{c}-Moody levels of $G_{\mathrm{SM}}$, we can,
from a bottom-up perspective, calculate the $U(1)_X$ gauge coupling
constant $g_X$ in those cases, where the $\Delta_i^{\ol{N}}$ are
uniquely fixed, \ie for all models with $k_C=2$. Evaluating the second
equality of Eq.~(\ref{ooooo}) yields values within the interval
\beq
g_X ~\in ~[0.0085\,,\,0.0145] \, ,
\eeq
which in turn enables us to calculate the mass of the heavy $U(1)_X$ vector
boson $B'$
\beq
m_{B'} ~\sim~ {g_X} \cdot \eps \cdot M_{\mathrm{grav}}~\approx ~ 5 \times 10^{15}
\, \mathrm{GeV}. 
\eeq
The results for each of the 6+24 models with uniquely fixed $X$-charge
assignments are listed in Tables~\ref{48details}+\ref{24details}.  We
point out that the $k_X$ corresponding to the above determined $g_X$
are quite high integers, \eg 8839 for $\#$~6 of Case~II. This
underlines that the scenarios without $X$-charged exotic matter are to
be taken more as an existence proof rather than concrete models.





\cleqn
\section{\label{conclusionsect}Discussion and 
Conclusion}

In this note, we have devised FN models in which the anomalous $U(1)_X
$ gauge symmetry is broken down to the discrete $\mathbb{Z}_6
$-symmetry, proton hexality. The masses of the light neutrino states
are generated by introducing right-handed neutrinos $\ol{N^i}$ and
applying the see-saw mechanism.  For Case~I, the Majorana mass terms
of $\ol{N^i}$ originate only from the FN-mechanism, while for Case~II
they result effectively from a combination of the FN- and the
GM/KN-mechanism. Requiring phenomenologically acceptable fermion
masses and mixings, the GS mixed anomaly cancellation conditions with
gauge coupling unification, as well as the low-energy remnant discrete
symmetry $P_6$, we are led to 48 $X$-charge assignments for Case~I
({\it cf.}  Table~\ref{parametersI}) and 504 $X$-charge assignments
for Case~II ({\it cf.}  Table~\ref{parametersII}).

Under the assumption of no exotic $X$-charged particles, all 48 sets of
Case~I, but only 24 of the 504 sets of Case~II are compatible with the GS
anomaly cancellation conditions. The $X$-charges of the resulting 48+24 sets
are shown in Tables~\ref{48sets} and \ref{24sets}.
Furthermore, we can determine the Ka\v{c}-Moody levels of $G_{\mathrm{SM}}$ in
these models. For $k_C=2$, the $X$-charges of the right-handed neutrinos are
fixed uniquely. This enables us to calculate the gauge coupling constant $g_X$
of $U(1)_X$ in these cases. 

All results are listed in Tables~\ref{48details} and \ref{24details}
together with the obtained light neutrino mass spectrum, the maximal
denominator of the $X$-charges, as well as some additional comments on
each of the models. We emphasize here that all are phenomenologically
acceptable because the unknown $\mathcal{O}(1)$ coefficients allow a
certain flexibility. However, if asked to select ``preferred'' models,
one can consider the following three criteria:
\begin{itemize}
\item[(1)]  ``nice'' CKM matrix,
\item[(2)]  naturally small CHOOZ mixing angle,
\item[(3)]  small maximal denominator for the $X$-charges.
\end{itemize}
Sets with $y=0$ lead to our preferred $\eps$-structure of the CKM
matrix, see Sect.~\ref{nonusect}. These amount to one third of all the
models. The CHOOZ mixing angle corresponds to the $(1,3)$ entry of the
MNS matrix. This is naturally suppressed in our models if $\Delta^L_
{31}=-2$ ($U^{\mathrm{MNS}}_{13} \sim \eps^2$) or $\Delta^L_{31}=-1$
($U^{\mathrm{MNS}}_{13} \sim \eps$), see the end of
Sect.~\ref{mixingsubsect}. Altogether 24+21 sets lead to a naturally
small CHOOZ angle by virtue of the $U(1)_X$ charge assignments.
Finally, we have labeled the 10+3 models with a maximal denominator
$\leq 54$ by ``denom.'' in the comments. From the aesthetical
viewpoint, the most appealing set is $\#$~6 of Case~II
(Table~\ref{24details}) where all $X$-charges are multiples of~$1/6$.
This model features a small CHOOZ angle but, unfortunately, a not so
nice CKM matrix. With regard to criterion~(3), we however emphasize
that models with highly-fractional $X$-charges are very common,
especially when fulfilling phenomenological constraints, see
Ref.~\cite{Dreiner:2003hw}.

Looking for models which satisfy all of the above three criteria, we find that
--~remarkably enough~--  only one remains: namely $\#$~32 of Case~I
(Table~\ref{48details}). This model has a normal hierarchical neutrino mass
spectrum with $U^{\mathrm{MNS}}_{13} \sim \eps$, the maximal denominator of
the $X$-charges is~30. Without $X$-charged hidden sector matter, $k_C=3$ and
$\sum_i \Delta_i^{\ol{N}} =18$, leading to 16 distinct $X$-charge assignments
for the right-handed neutrinos, {\it cf.} Appendix~\ref{tablesect}.





\section*{Acknowledgments}

We are grateful for helpful discussions with or useful comments from
Alon Faraggi, St\'ephane Lavignac, Thomas Mohaupt, Pi\`erre Ramond,
Carlos Savoy and Martin Walter. C.L. thanks the SPhT at the
CEA-Saclay, and C.L. as well as M.T. thank the Physikalisches Institut
in Bonn for hospitality. This project is partially supported by
the European Union 6th framework program MRTN-CT-2004-503369 ``Quest
for unification". The work of H.D. is also partially supported 
by the RTN European program MRTN-CT-2004-005104 ``ForcesUniverse",
MRTN-CT-2006-035863 ``UniverseNet" and by the SFB-Transregio~33 ``The Dark
Universe" of the Deutsche Forschungsgemeinschaft (DFG). M.T. greatly
appreciates that he was funded by a Feodor Lynen fellowship of the
Alexander-von-Humboldt-Foundation. The work of H.M. is supported in part by
the DOE under the contract DE-AC03-76SF00098 and in part by NSF grant
PHY-0098840.





\begin{appendix}

\cleqn
~\\
\\
\noindent{\LARGE \bf Appendix}


\section{\label{detailsect}$\boldsymbol{X_{\ol{N^a}}}\:
$-Dependence of the
  Neutrino Masses}

For Case~II [{\it cf.} Eq.~(\ref{massII})], the Dirac and
the Majorana mass matrices can be written as
\beq
M^{(\mathrm{D})}_{ij} ~=~ A \cdot \alpha_{ij} \:  \eps^{X_{L^i}
  + X_{\ol{N^j}}}, ~~~~~~~~~
M^{(\mathrm{M})}_{ij} ~=~ B \cdot \beta_{ij} \:  \eps^{-X_{\ol{N^i}} -X_{\ol{N^j}}},
\eeq
with $A\equiv \langle H^U \rangle \, \eps^{X_{H^U}}$ and $B\equiv m_{3/2}$. 
The dimensionless coefficients $\alpha_{ij}$ and $\beta_{ij}$
are of order one. In our basis, $M^{(\mathrm{M})}_{ij}$ and thus 
$\beta_{ij}$ is diagonal. With this notation the effective light neutrino
mass matrix reads
\beqn
M^{(\nu,\mathrm{II})}_{ij} &=& -\,\frac{A^2}{B} \cdot \sum_k
 \frac{\alpha_{ik} \: \alpha_{jk}}{\beta_{kk}} \: 
 \eps^{X_{L^i} + X_{L^j} + 4X_{\ol{N^k}}} \notag \\
&=&  -\,\frac{A^2}{B} \cdot \sum_k a_{ik} \: a_{jk}.
\eeqn
In the last step we have defined $a_{ik} \equiv
\frac{\alpha_{ik}}{\sqrt{\beta_{kk}}} \: \eps^{X_{L^i} + 2X_{\ol{N^k}}} $. 
The light neutrino masses $\widetilde{m} = \frac{A^2}{B} \lambda$ can now 
be obtained from the characteristic
polynomial\footnote{$\boldsymbol{M^{(\nu)}}$ can be diagonalized by a unitary
  matrix   $\boldsymbol{V}$. From  $\boldsymbol{V^T \cdot M^{(\nu)} \cdot V} =
\boldsymbol{M^{(\nu)}_{\mathrm{diag}}} $ we obtain the equation
$\boldsymbol{M^{(\nu)}} \vec{v} = \widetilde{m}\,
\vec{v}^\ast = \widetilde{m} \, \boldsymbol{P} \vec{v}$. Here, $\vec{v}$ is
one of the three normalized vectors of $\boldsymbol{V}$, and $\boldsymbol{P}$
is a diagonal matrix with $P_{ii} =  p_i = \frac{v^\ast_i}{v_i}$. The singular
values $\widetilde{m}$ of $\boldsymbol{M^{(\nu)}}$ are determined by the
condition $\mathrm{det} ( \boldsymbol{M^{(\nu)}} - \widetilde{m} \,
\boldsymbol{P} ) = 0$, which -- up to the phase factors $p_i$ -- is just the
characteristic polynomial.}
of $\boldsymbol{M^{(\nu,\mathrm{II})}}$,
\beq
C_3\lambda^3 + C_2 \lambda^2 + C_1 \lambda + C_0 ~=~0,\label{characteristic}
\eeq
where
\beqn
C_3&=&p_1p_2p_3,
\eeqn
\beqn
C_2 &=& 
p_1 p_2 (a_{33}^2 + a_{32}^2 + a_{31}^2) 
+ p_1 p_3 (a_{23}^2 + a_{22}^2 + a_{21}^2) 
+ p_2 p_3 (a_{13}^2 + a_{12}^2 + a_{11}^2),~~~
\eeqn
\beqn
C_1 &=& \;\:\;
\,p_1 \left[(a_{33}a_{22} - a_{32}a_{23})^2 
+ (a_{31}a_{23} - a_{33}a_{21})^2 
+ (a_{32}a_{21} - a_{31}a_{22})^2\right] \notag \\
&& + \,p_2 \left[ (a_{13}a_{32} - a_{12}a_{33})^2 
+ (a_{11}a_{33} - a_{13}a_{31})^2 
+ (a_{12}a_{31} - a_{11}a_{32})^2\right] \notag \\
&& + \,p_3 \left[(a_{23}a_{12} - a_{22}a_{13})^2 
+ (a_{21}a_{13} - a_{23}a_{11})^2 
+ (a_{22}a_{11} - a_{21}a_{12})^2 \right],~
\eeqn
\beqn
C_0~=~ (a_{33}a_{22}a_{11} + a_{31}a_{23}a_{12} + a_{32}a_{21}a_{13}
 - a_{33}a_{21}a_{12} - a_{31}a_{22}a_{13} - a_{32}a_{23}a_{11})^2.~~~~~~
\eeqn
As $a_{ik}\sim \eps^{X_{L^i} + 2 X_{\ol{N^k}}}$, the {\it order} of the
coefficients $C_{...}$ can be readily determined. With 
$X_{L^3} \leq X_{L^2} \leq X_{L^1}$ and 
$X_{\ol{N^3}} \leq X_{\ol{N^2}} \leq X_{\ol{N^1}}$ we get
\beqn
C_3 & = & c_3,\\
C_2 & = & c_2 \: \eps^{2X_{L^3}+4X_{\ol{N^3}}} ,\\
C_1 & = & c_1 \: \eps^{2X_{L^2} + 2X_{L^3} + 4X_{\ol{N^2}} +4X_{\ol{N^3}}} ,\\
C_0 & = & c_0 \: \eps^{2X_{L^1}+2X_{L^2}+2X_{L^3}+ 
4X_{\ol{N^1}}+ 4X_{\ol{N^2}}+ 4X_{\ol{N^3}}},
\eeqn
where $c_3,c_2,c_1,c_0$ are $\mathcal{O}(1)$ coefficients.  Inserting
these expressions into Eq.~(\ref{characteristic}), the three singular
values $\lambda$ can be obtained. The order of the largest $\lambda$
depends only on the cubic and the quadratic term: Assuming [this is
justified in hindsight\footnote{This method is akin to the slow roll
  approximation in inflationary cosmology: The Klein-Gordon equation
  for a homogeneous scalar field $\varphi$ reads
  $\ddot{\varphi}+3H\dot{\varphi}+m^2\varphi=0$, $H$ being the Hubble
  parameter. Assuming slow roll, \textit{i.e.} $\ddot{\varphi} \ll
  \{H\dot{\varphi},m^2\varphi\}$, yields
  $3H\dot{\varphi}+m^2\varphi=0$, which's solution in hindsight
  justifies the slow roll approximation.}  from the result
Eq.~(\ref{lamdrei})]
\beq
C_3\lambda^3,C_2\lambda^2 > C_1\lambda,C_0,
\eeq 
we get $C_3\lambda+C_2=0$, which yields
\beq\label{lamdrei}
\lambda_3 ~=~ - \frac{c_2}{c_3}  \: \eps^{2X_{L^3}+4X_{\ol{N^3}}} 
+ \mathrm{equal/higher~orders},
\eeq
where ``equal'' applies only if $X_{L^2} = X_{L^3}$ and $X_{\ol{N^2}} =
X_{\ol{N^3}}$. 
Similarly, the order of the second singular 
value is derived from the quadratic and the linear
term of Eq.~(\ref{characteristic})
\beq
\lambda_2 ~=~ - \frac{c_1}{c_2}  \: \eps^{2X_{L^2}+4X_{\ol{N^2}}} +
\mathrm{equal/higher~orders},
\eeq
where ``equal'' applies only if  either $X_{L^2} = X_{L^3}$ and $X_{\ol{N^2}} =
X_{\ol{N^3}}$ or   $X_{L^1} = X_{L^2}$ and $X_{\ol{N^1}} = X_{\ol{N^2}}$.
Finally, the order of
$\lambda_1$ is obtained from the linear and the constant term
\beq
\lambda_1 ~=~ - \frac{c_0}{c_1}  \: \eps^{2X_{L^1}+4X_{\ol{N^1}}} +
\mathrm{equal/higher~orders},
\eeq
where ``equal'' applies only if $X_{L^1} = X_{L^2}$ and $X_{\ol{N^1}} =
X_{\ol{N^2}}$. 
This yields the following ratios for the light neutrino masses
\beq
\widetilde{m}_3 ~:~ \widetilde{m}_2 ~:~ \widetilde{m}_1 ~~\sim~~
\eps^{2 X_{L^3} + 4X_{\ol{N^3}}} ~:~ \eps^{2X_{L^2}+4X_{\ol{N^2}}} ~:~
      \eps^{2X_{L^1} +4X_{\ol{N^1}}}.\label{rightXII}
\eeq
Analogously, we obtain for Case~III that 
\beq
\widetilde{m}_1 ~:~ \widetilde{m}_2 ~:~ \widetilde{m}_3 ~~\sim~~
\eps^{-2X_{L^1} -4X_{\ol{N^1}}} ~:~ \eps^{-2X_{L^2}-4X_{\ol{N^2}}} ~:~
      \eps^{-2 X_{L^3}-4X_{\ol{N^3}}}. \label{rightXIII}
\eeq





\cleqn
\section{\label{tablesect}Tables of 
$\boldsymbol{X}$-Charges}
Combining Table~\ref{parametersI} with Table~\ref{Table2} leads to the
$X$-charge assignments of Table~\ref{48sets} (Case~I): 
\eg the choice $\Delta_{31}^L=0$, $3\zeta+p=-1$ and $\Delta ^H=-3$,
with $z=1$, yields for instance 
$$
X_{H^D}~=~\frac{4 x \cdot (3 x+28)+36 y+193}{30 (x+7)}.
$$
Then one picks a value for $x$ and a value for $y$. Table~\ref{48details}
displays some features of these 48 possibilities.
In search for a low fractionality for the $X$-charges 
one finds with $y=0$ cases $\#\,$8 (54),
$\#\,$17 (48), $\#\,$32 (30), with $y=1$ cases  $\#\,$21 (30), $\#\,$27 (42),
with $y=-1$ cases $\#\,$19 (18), $\#\,$22 (12), $\#\,$25 (30), $\#\,$31 (18),
$\#\,$43 (30).  The numbers in parentheses give the maximal denominators of the
$X$-charges, in the normalization where $X_A=-1$.

Assuming no $X$-charged hidden sector superfields, one can determine the
Ka\v{c}-Moody levels $k_C$ and the {\it sum} of the $\Delta_i^{\ol{N}}$.
Recalling the constraints of Eqs.~(\ref{NhierI},\ref{NdegI}), we find for
$k_C=3$ that $(\Delta_1^{\ol{N}}, \Delta_2^{\ol{N}}, \Delta_3^{\ol{N}})$ can
take the following values, respectively:
\begin{itemize}
\item[]$\hspace{-10mm}\sum_i\Delta_i^{\ol{N}}=23$:
$(8, 8, 7)$,  $(9, 7, 7)$,  $(9, 8, 6)$,  $(9, 9, 5)$, $(10, 7, 6)$, 
 $(10, 8, 5)$,  $(10, 9, 4)$, \\ $~\hspace{14mm}$
 $(10, 10, 3)$,  $(11, 6, 6)$,  $(11, 7, 5)$, $(11, 8, 4)$, 
 $(11, 9, 3)$, $(12, 6, 5)$,  $(12, 7, 4)$,
\item[]$\hspace{-10mm}\sum_i\Delta_i^{\ol{N}}=20$:    
$(7, 7, 6)$, $(8, 6, 6)$, $(8, 7, 5)$, $(8, 8, 4)$, $(9, 6, 5)$, $(9, 7, 4)$,
$(9, 8, 3)$, $(9, 9, 2)$,  \\ $~\hspace{14mm}$ $(10, 5, 5)$, $(10, 6, 4)$,
$(10, 7, 3)$, $(10, 8, 2)$, $(11, 5, 4)$, $(11, 6, 3)$, $(11, 7, 2)$, 
\\ $~\hspace{14mm}$ $(12, 4, 4)$, $(12, 5, 3)$,
\item[]$\hspace{-10mm}\sum_i\Delta_i^{\ol{N}}=18$:
$(6, 6, 6)$, $(7, 6, 5)$, $(7, 7, 4)$, $(8, 5, 5)$, $(8, 6, 4)$, $(8, 7, 3)$,
$(8, 8, 2)$, $(9, 5, 4)$,\\ $~\hspace{14mm}$ $(9, 6, 3)$, $(9, 7, 2)$, 
$(10,4,4)$, $(10,5,3)$, $(10, 6,2)$, $(11, 4, 3)$, $(11, 5, 2)$, $(12, 3, 3)$,
\item[]$\hspace{-10mm}\sum_i\Delta_i^{\ol{N}}=17$: 
$(6, 6, 5)$, $(7, 5, 5)$, $(7, 6, 4)$, $(7, 7, 3)$, $(8, 5, 4)$, $(8, 6, 3)$,
$(8, 7, 2)$, \\  $~\hspace{14mm}$ $(9, 4, 4)$, $(9, 5, 3)$, $(9, 6, 2)$, 
$(10, 4, 3)$, $(10, 5,2)$, $(11, 3, 3)$, $(11, 4, 2)$.
\end{itemize}
For $k_C=2$, the $\Delta_i^{\ol{N}}$ are uniquely fixed and given in
Table~\ref{48details}.

  It is interesting to note that the lower limit from thermal
  leptogenesis, $M^{(\mathrm{M})}_{11} \gtrsim 4 \times 10^8$~GeV
  \cite{Buchmuller:2002jk}, requires $1 + 2 \Delta_1^{\ol{N}} \lesssim 15$,
  and hence $\Delta_1^{\ol{N}} \lesssim 7$.  We observe that there is only a
  small number of 
  combinations allowed within this limit [{\it e.g.} for $\sum_i
  \Delta_i^{\overline{N}} = 20$ only $(7,7,6)$ is okay]. 
  On the other hand, some of
  the solutions above predict no hierarchy between $\overline{N^1}$ and
  $\overline{N^2}$, and the bound may be less severe, {\it e.g.} $2 \times
  10^7$~GeV in Ref.~\cite{Giudice:2003jh}. In the extreme case of resonant
  enhancement, one can allow for even TeV scale right-handed neutrinos
  \cite{Pilaftsis:2003gt}.

Case~II is treated similarly. However, displaying explicitly the 504
sets of $X$-charges which are hinted at in Table~\ref{parametersII}
would fill more than 12~pages.  We content ourselves with presenting
those 24~models which are consistent without $X$-charged exotic
matter. They are given in Tables~\ref{24sets} and \ref{24details}.
Small maximal denominators of the $X$-charges are obtained for cases
$\#\,$6 (6), $\#\,$7 (30), $\#\,$9 (42).

\begin{table}
\begin{center}
\tiny
$~\hspace{-3mm}
\begin{array}{|r||r|r|r|r|r|r|r|r|r|r|r|r|r|r|r|r|r|}
\hline
 \phantom{\bigg|}\# & X_{H^D} & X_{H^U} & X_{Q^1} & X_{Q^2} & X_{Q^3} & X_{\overline{U^1}} & X_{\overline{U^2}} & X_{\overline{U^3}} & X_{\overline{D^1}} & X_{\overline{D^2}} & X_{\overline{D^3}} & 
X_{L^1} & X_{L^2} & X_{L^3} & X_{\overline{E^1}} & X_{\overline{E^2}} & X_{\overline{E^3}} \\
\hline
\hline\phantom{\Big|}  1 & \frac{157}{210} & -  \frac{367}{210}    & \frac{388}{105} & \frac{388}{105} & \frac{178}{105} & \frac{1271}{210} & \frac{431}{210} & \frac{11}{210} & - \frac{31}{70}    & -  \frac{171}{70}    & -  \frac{171}{70}    & -  \frac{184}{105}    & -  \frac{184}{105}& -  \frac{184}{105}    & \frac{1261}{210} & \frac{631}{210} & \frac{211}{210} \\
\hline\phantom{\Big|}  2 & \frac{193}{210} & -  \frac{403}{210} & \frac{452}{105} & \frac{347}{105} & \frac{137}{105} & \frac{393}{70} & \frac{183}{70} & \frac{43}{70} & -  \frac{257}{210}    & -  \frac{467}{210}    & -  \frac{467}{210}    & -  \frac{166}{105}    & -  \frac{166}{105}    & -  \frac{166}{105}    & \frac{1189}{210} & \frac{559}{210} & \frac{139}{210} \\
\hline\phantom{\Big|}  3 & \frac{229}{210} & -  \frac{439}{210}    & \frac{172}{35} & \frac{102}{35} & \frac{32}{35} & \frac{1087}{210} & \frac{667}{210} & \frac{247}{210} & -  \frac{421}{210}    & -  \frac{421}{210}  & -  \frac{421}{210}    & -  \frac{148}{105}    & -  \frac{148}{105}    & -  \frac{148}{105}    & \frac{1117}{210} & \frac{487}{210} & \frac{67}{210} \\
\hline\phantom{\Big|}  4 & \frac{281}{240} & -  \frac{521}{240}    & \frac{311}{80} & \frac{311}{80} & \frac{151}{80} & \frac{377}{60} & \frac{137}{60} & \frac{17}{60} & -  \frac{7}{120}    & -  \frac{247}{120}    & -  \frac{247}{120} &  -\frac{319}{240}    & -  \frac{319}{240}    & -  \frac{319}{240}    & \frac{739}{120} & \frac{379}{120} & \frac{139}{120}\\
\hline\phantom{\Big|}  5 & \frac{317}{240} & -  \frac{557}{240}    & \frac{1081}{240} & \frac{841}{240} & \frac{361}{240} & \frac{349}{60} & \frac{169}{60} &\frac{49}{60} & -  \frac{33}{40}    & -  \frac{73}{40}    & -  \frac{73}{40}   & -  \frac{283}{240}  & -  \frac{283}{240}    & -  \frac{283}{240}    & \frac{703}{120} & \frac{343}{120} & \frac{103}{120} \\
\hline\phantom{\Big|}  6 & \frac{353}{240} & -\frac{593}{240}    & \frac{1229}{240} & \frac{749}{240} & \frac{269}{240} & \frac{107}{20} & \frac{67}{20} & \frac{27}{20} & -  \frac{191}{120}    & -  \frac{191}{120}    & -  \frac{191}{120}    & -  \frac{247}{240}    & -  \frac{247}{240}  & -  \frac{247}{240}    & \frac{667}{120} & \frac{307}{120} & \frac{67}{120}\\ 
\hline\phantom{\Big|}  7 & \frac{143}{90} & -  \frac{233}{90}    & \frac{551}{135} & \frac{551}{135} & \frac{281}{135} & \frac{1757}{270} & \frac{677}{270} & \frac{137}{270} & \frac{89}{270} & -  \frac{451}{270}  & -  \frac{451}{270}    & -  \frac{41}{45}    & -  \frac{41}{45}    & -  \frac{41}{45}    & \frac{569}{90} & \frac{299}{90} & \frac{119}{90}\\
\hline\phantom{\Big|}  8 & \frac{31}{18} & -  \frac{49}{18}    & \frac{127}{27} & \frac{100}{27} & \frac{46}{27} & \frac{325}
   {54} & \frac{163}{54} & \frac{55}{54} & -  \frac{23}{54}    & -  \frac{77}{54}    & -  \frac{77}{54}    & -  \frac{7}
     {9}    & -  \frac{7}{9}    & -  \frac{7}{9}    & \frac{109}{18} & \frac{55}{18} & \frac{19}{18} \\ \hline\phantom{\Big|}  9 & \frac{167}{90} & - 
     \frac{257}{90}    & \frac{719}{135} & \frac{449}{135} & \frac{179}{135} & \frac{1493}{270} & \frac{953}{270} & \frac{413}{270} & -  \frac{319}
     {270}    & -  \frac{319}{270}    & -  \frac{319}{270}    & -  \frac{29}{45}    & -  \frac{29}{45}  
      & -  \frac{29}{45}    & \frac{521}{90} & \frac{251}{90} & \frac{71}{90} \\ \hline\phantom{\Big|}  10 & \frac{601}{300} & -  \frac{901}{300}    & \frac{1283}
   {300} & \frac{1283}{300} & \frac{683}{300} & \frac{1009}{150} & \frac{409}{150} & \frac{109}{150} & \frac{18}{25} & -  \frac{32}{25}    & - 
     \frac{32}{25}    & -  \frac{149}{300}    & -  \frac{149}{300}    & -  \frac{149}{300}    & \frac{487}{75} & \frac{262}
   {75} & \frac{112}{75} \\ \hline\phantom{\Big|}  11 & \frac{637}{300} & -  \frac{937}{300}    & \frac{1471}{300} & \frac{1171}{300} & \frac{571}{300} & \frac{311}
   {50} & \frac{161}{50} & \frac{61}{50} & -  \frac{2}{75}    & -  \frac{77}{75}    & -  \frac{77}{75}    & -  \frac{113}
     {300}    & -  \frac{113}{300}    & -  \frac{113}{300}    & \frac{469}{75} & \frac{244}{75} & \frac{94}{75} \\ \hline\phantom{\Big|}  12 & \frac{673}
   {300} & -  \frac{973}{300}    & \frac{553}{100} & \frac{353}{100} & \frac{153}{100} & \frac{857}{150} & \frac{557}{150} & \frac{257}{150} & - 
     \frac{58}{75}    & -  \frac{58}{75}    & -  \frac{58}{75}    & -  \frac{77}{300}    & -  \frac{77}{300}  
      & -  \frac{77}{300}    & \frac{451}{75} & \frac{226}{75} & \frac{76}{75} \\ \hline\phantom{\Big|}  13 & \frac{67}{210} & -  \frac{277}{210}    & \frac{368}
   {105} & \frac{368}{105} & \frac{158}{105} & \frac{407}{70} & \frac{127}{70} & -  \frac{13}{70}    & \frac{37}{210} & -  \frac{383}{210}  
      & -  \frac{383}{210}    & -  \frac{124}{105}    & -  \frac{124}{105}    & -  \frac{124}{105}    & \frac{1231}
   {210} & \frac{601}{210} & \frac{181}{210} \\ \hline\phantom{\Big|}  14 & \frac{103}{210} & -  \frac{313}{210}    & \frac{144}{35} & \frac{109}{35} & \frac{39}{35} & 
    \frac{1129}{210} & \frac{499}{210} & \frac{79}{210} & -  \frac{127}{210}    & -  \frac{337}{210}    & -  \frac{337}{210}  
      & -  \frac{106}{105}    & -  \frac{106}{105}    & -  \frac{106}{105}    & \frac{1159}{210} & \frac{529}{210} & \frac{109}
   {210} \\ \hline\phantom{\Big|}  15 & \frac{139}{210} & -  \frac{349}{210}    & \frac{496}{105} & \frac{286}{105} & \frac{76}{105} & \frac{1037}{210} & \frac{617}
   {210} & \frac{197}{210} & -  \frac{97}{70}    & -  \frac{97}{70}    & -  \frac{97}{70}    & -  \frac{88}{105}  
      & -  \frac{88}{105}    & -  \frac{88}{105}    & \frac{1087}{210} & \frac{457}{210} & \frac{37}{210} \\ \hline\phantom{\Big|}  16 & \frac{179}{240} & - 
     \frac{419}{240}    & \frac{887}{240} & \frac{887}{240} & \frac{407}{240} & \frac{121}{20} & \frac{41}{20} & \frac{1}{20} & \frac{67}{120} & - 
     \frac{173}{120}    & -  \frac{173}{120}    & -  \frac{181}{240}    & -  \frac{181}{240}    & -  \frac{181}
     {240}    & \frac{721}{120} & \frac{361}{120} & \frac{121}{120} \\ \hline\phantom{\Big|}  17 & \frac{43}{48} & -  \frac{91}{48}    & \frac{69}{16} & \frac{53}
   {16} & \frac{21}{16} & \frac{67}{12} & \frac{31}{12} & \frac{7}{12} & -  \frac{5}{24}    & -  \frac{29}{24}    & -  \frac{29}
     {24}    & -  \frac{29}{48}    & -  \frac{29}{48}    & -  \frac{29}{48}    & \frac{137}{24} & \frac{65}{24} & \frac{17}
   {24} \\ \hline\phantom{\Big|}  18 & \frac{251}{240} & -  \frac{491}{240}    & \frac{1183}{240} & \frac{703}{240} & \frac{223}{240} & \frac{307}{60} & \frac{187}
   {60} & \frac{67}{60} & -  \frac{39}{40}    & -  \frac{39}{40}    & -  \frac{39}{40}    & -  \frac{109}{240}  
      & -  \frac{109}{240}    & -  \frac{109}{240}    & \frac{649}{120} & \frac{289}{120} & \frac{49}{120} \\ \hline\phantom{\Big|}  19 & \frac{7}{6} & - 
     \frac{13}{6}    & \frac{35}{9} & \frac{35}{9} & \frac{17}{9} & \frac{113}{18} & \frac{41}{18} & \frac{5}{18} & \frac{17}{18} & -  \frac{19}
     {18}    & -  \frac{19}{18}    & -  \frac{1}{3}    & -  \frac{1}{3}    & -  \frac{1}{3}    & \frac{37}
   {6} & \frac{19}{6} & \frac{7}{6} \\ \hline\phantom{\Big|}  20 & \frac{13}{10} & -  \frac{23}{10}    & \frac{203}{45} & \frac{158}{45} & \frac{68}{45} & \frac{521}
   {90} & \frac{251}{90} & \frac{71}{90} & \frac{17}{90} & -  \frac{73}{90}    & -  \frac{73}{90}    & -  \frac{1}{5}    & -
       \frac{1}{5}    & -  \frac{1}{5}    & \frac{59}{10} & \frac{29}{10} & \frac{9}{10} \\ \hline\phantom{\Big|}  21 & \frac{43}{30} & -  \frac{73}{30}  
      & \frac{77}{15} & \frac{47}{15} & \frac{17}{15} & \frac{53}{10} & \frac{33}{10} & \frac{13}{10} & -  \frac{17}{30}    & -  \frac{17}
     {30}    & -  \frac{17}{30}    & -  \frac{1}{15}    & -  \frac{1}{15}    & -  \frac{1}{15}    & \frac{169}
   {30} & \frac{79}{30} & \frac{19}{30} \\ \hline\phantom{\Big|}  22 & \frac{19}{12} & -  \frac{31}{12}    & \frac{49}{12} & \frac{49}{12} & \frac{25}{12} & \frac{13}
   {2} & \frac{5}{2} & \frac{1}{2} & \frac{4}{3} & -  \frac{2}{3}    & -  \frac{2}{3}    & \frac{1}{12} & \frac{1}{12} & \frac{1}
   {12} & \frac{19}{3} & \frac{10}{3} & \frac{4}{3} \\ \hline\phantom{\Big|}  23 & \frac{511}{300} & -  \frac{811}{300}    & \frac{471}{100} & \frac{371}{100} & \frac{171}
   {100} & \frac{899}{150} & \frac{449}{150} & \frac{149}{150} & \frac{44}{75} & -  \frac{31}{75}    & -  \frac{31}{75}    & \frac{61}
   {300} & \frac{61}{300} & \frac{61}{300} & \frac{457}{75} & \frac{232}{75} & \frac{82}{75} \\ \hline\phantom{\Big|}  24 & \frac{547}{300} & -  \frac{847}{300}    & 
    \frac{1601}{300} & \frac{1001}{300} & \frac{401}{300} & \frac{823}{150} & \frac{523}{150} & \frac{223}{150} & -  \frac{4}{25}    & -  \frac{4}
     {25}    & -  \frac{4}{25}    & \frac{97}{300} & \frac{97}{300} & \frac{97}{300} & \frac{439}{75} & \frac{214}{75} & \frac{64}
   {75} \\ \hline\phantom{\Big|}  25 & -  \frac{1}{30}    & -  \frac{29}{30}    & \frac{17}{5} & \frac{17}{5} & \frac{7}{5} & \frac{167}{30} & \frac{47}
   {30} & -  \frac{13}{30}    & \frac{19}{30} & -  \frac{41}{30}    & -  \frac{41}{30}    & -  \frac{8}{15}    & 
    -  \frac{8}{15}    & -  \frac{23}{15}    & \frac{167}{30} & \frac{77}{30} & \frac{47}{30} \\ \hline\phantom{\Big|}  26 & \frac{29}{210} & -  \frac{239}
     {210}    & \frac{421}{105} & \frac{316}{105} & \frac{106}{105} & \frac{359}{70} & \frac{149}{70} & \frac{9}{70} & -  \frac{31}{210}  
      & -  \frac{241}{210}    & -  \frac{241}{210}    & -  \frac{38}{105}    & -  \frac{38}{105}    & -  
      \frac{143}{105}    & \frac{1097}{210} & \frac{467}{210} & \frac{257}{210} \\ \hline\phantom{\Big|}  27 & \frac{13}{42} & -  \frac{55}{42}    & \frac{97}
   {21} & \frac{55}{21} & \frac{13}{21} & \frac{197}{42} & \frac{113}{42} & \frac{29}{42} & -  \frac{13}{14}    & -  \frac{13}{14}  
      & -  \frac{13}{14}    & -  \frac{4}{21}    & -  \frac{4}{21}    & -  \frac{25}{21}    & \frac{205}{42} & 
    \frac{79}{42} & \frac{37}{42} \\ \hline\phantom{\Big|}  28 & \frac{97}{240} & -  \frac{337}{240}    & \frac{287}{80} & \frac{287}{80} & \frac{127}{80} & \frac{349}
   {60} & \frac{109}{60} & -  \frac{11}{60}    & \frac{121}{120} & -  \frac{119}{120}    & -  \frac{119}{120}    & -  
      \frac{23}{240}    & -  \frac{23}{240}    & -  \frac{263}{240}    & \frac{683}{120} & \frac{323}{120} & \frac{203}
   {120} \\ \hline\phantom{\Big|}  29 & \frac{133}{240} & -  \frac{373}{240}    & \frac{1009}{240} & \frac{769}{240} & \frac{289}{240} & \frac{107}{20} & \frac{47}
   {20} & \frac{7}{20} & \frac{29}{120} & -  \frac{91}{120}    & -  \frac{91}{120}    & \frac{13}{240} & \frac{13}{240} & -  \frac{227}
     {240}    & \frac{647}{120} & \frac{287}{120} & \frac{167}{120} \\ \hline\phantom{\Big|}  30 & \frac{169}{240} & -  \frac{409}{240}    & \frac{1157}{240} & 
    \frac{677}{240} & \frac{197}{240} & \frac{293}{60} & \frac{173}{60} & \frac{53}{60} & -  \frac{21}{40}    & -  \frac{21}{40}    & -
       \frac{21}{40}    & \frac{49}{240} & \frac{49}{240} & -  \frac{191}{240}    & \frac{611}{120} & \frac{251}{120} & \frac{131}
   {120} \\ \hline\phantom{\Big|}  31 & \frac{5}{6} & -  \frac{11}{6}    & \frac{34}{9} & \frac{34}{9} & \frac{16}{9} & \frac{109}{18} & \frac{37}{18} & \frac{1}{18} & 
    \frac{25}{18} & -  \frac{11}{18}    & -  \frac{11}{18}    & \frac{1}{3} & \frac{1}{3} & -  \frac{2}{3}    & \frac{35}
   {6} & \frac{17}{6} & \frac{11}{6} \\ \hline\phantom{\Big|}  32 & \frac{29}{30} & -  \frac{59}{30}    & \frac{22}{5} & \frac{17}{5} & \frac{7}{5} & \frac{167}{30} & 
    \frac{77}{30} & \frac{17}{30} & \frac{19}{30} & -  \frac{11}{30}    & -  \frac{11}{30}    & \frac{7}{15} & \frac{7}{15} & -  
      \frac{8}{15}    & \frac{167}{30} & \frac{77}{30} & \frac{47}{30} \\ \hline\phantom{\Big|}  33 & \frac{11}{10} & -  \frac{21}{10}    & \frac{226}{45} & \frac{136}
   {45} & \frac{46}{45} & \frac{457}{90} & \frac{277}{90} & \frac{97}{90} & -  \frac{11}{90}    & -  \frac{11}{90}    & -  \frac{11}
     {90}    & \frac{3}{5} & \frac{3}{5} & -  \frac{2}{5}    & \frac{53}{10} & \frac{23}{10} & \frac{13}{10} \\ \hline\phantom{\Big|}  34 & \frac{377}{300} & - 
     \frac{677}{300}    & \frac{397}{100} & \frac{397}{100} & \frac{197}{100} & \frac{943}{150} & \frac{343}{150} & \frac{43}{150} & \frac{133}{75} & -
       \frac{17}{75}    & -  \frac{17}{75}    & \frac{227}{300} & \frac{227}{300} & -  \frac{73}{300}    & \frac{449}{75} & \frac{224}
   {75} & \frac{149}{75} \\ \hline\phantom{\Big|}  35 & \frac{413}{300} & -  \frac{713}{300}    & \frac{1379}{300} & \frac{1079}{300} & \frac{479}{300} & \frac{289}
   {50} & \frac{139}{50} & \frac{39}{50} & \frac{77}{75} & \frac{2}{75} & \frac{2}{75} & \frac{263}{300} & \frac{263}{300} & -  \frac{37}{300}  
      & \frac{431}{75} & \frac{206}{75} & \frac{131}{75} \\ \hline\phantom{\Big|}   36 & \frac{449}{300} & -  \frac{749}{300}    & \frac{1567}{300} & \frac{967}{300} & 
    \frac{367}{300} & \frac{791}{150} & \frac{491}{150} & \frac{191}{150} & \frac{7}{25} & \frac{7}{25} & \frac{7}{25} & \frac{299}{300} & \frac{299}{300} & 
 \frac{1}{300}  & \frac{413}{75} & \frac{188}{75} & \frac{113}{75} \\\hline\phantom{\Big|}            37 & -\frac{53}{210} & -\frac{157}{210} & \frac{353}{105} & \frac{353}{105} & \frac{143}{105} & \frac{377}{70} &
   \frac{97}{70} & -\frac{43}{70} & \frac{187}{210} & -\frac{233}{210} & -\frac{233}{210} & \frac{26}{105} & -\frac{79}{105}
   & -\frac{184}{105} & \frac{1051}{210} & \frac{631}{210} & \frac{421}{210} \\
  \hline\phantom{\Big|}  38 & -\frac{17}{210} & -\frac{193}{210} & \frac{139}{35} & \frac{104}{35} & \frac{34}{35} & \frac{1039}{210} &
   \frac{409}{210} & -\frac{11}{210} & \frac{23}{210} & -\frac{187}{210} & -\frac{187}{210} & \frac{44}{105} &
   -\frac{61}{105} & -\frac{166}{105} & \frac{979}{210} & \frac{559}{210} & \frac{349}{210} \\
  \hline\phantom{\Big|}  39 & \frac{19}{210} & -\frac{229}{210} & \frac{481}{105} & \frac{271}{105} & \frac{61}{105} & \frac{947}{210} &
   \frac{527}{210} & \frac{107}{210} & -\frac{47}{70} & -\frac{47}{70} & -\frac{47}{70} & \frac{62}{105} & -\frac{43}{105} &
   -\frac{148}{105} & \frac{907}{210} & \frac{487}{210} & \frac{277}{210} \\
  \hline\phantom{\Big|}  40 & \frac{47}{240} & -\frac{287}{240} & \frac{851}{240} & \frac{851}{240} & \frac{371}{240} & \frac{113}{20} &
   \frac{33}{20} & -\frac{7}{20} & \frac{151}{120} & -\frac{89}{120} & -\frac{89}{120} & \frac{167}{240} & -\frac{73}{240} &
   -\frac{313}{240} & \frac{613}{120} & \frac{373}{120} & \frac{253}{120} \\
  \hline\phantom{\Big|}  41 & \frac{83}{240} & -\frac{323}{240} & \frac{333}{80} & \frac{253}{80} & \frac{93}{80} & \frac{311}{60} & \frac{131}{60} &
   \frac{11}{60} & \frac{59}{120} & -\frac{61}{120} & -\frac{61}{120} & \frac{203}{240} & -\frac{37}{240} & -\frac{277}{240}
   & \frac{577}{120} & \frac{337}{120} & \frac{217}{120} \\
  \hline\phantom{\Big|}  42 & \frac{119}{240} & -\frac{359}{240} & \frac{1147}{240} & \frac{667}{240} & \frac{187}{240} & \frac{283}{60} &
   \frac{163}{60} & \frac{43}{60} & -\frac{11}{40} & -\frac{11}{40} & -\frac{11}{40} & \frac{239}{240} & -\frac{1}{240} &
   -\frac{241}{240} & \frac{541}{120} & \frac{301}{120} & \frac{181}{120} \\
  \hline\phantom{\Big|}  43 & \frac{19}{30} & -\frac{49}{30} & \frac{56}{15} & \frac{56}{15} & \frac{26}{15} & \frac{59}{10} & \frac{19}{10} &
   -\frac{1}{10} & \frac{49}{30} & -\frac{11}{30} & -\frac{11}{30} & \frac{17}{15} & \frac{2}{15} & -\frac{13}{15} &
   \frac{157}{30} & \frac{97}{30} & \frac{67}{30} \\
  \hline\phantom{\Big|}  44 & \frac{23}{30} & -\frac{53}{30} & \frac{196}{45} & \frac{151}{45} & \frac{61}{45} & \frac{487}{90} & \frac{217}{90} &
   \frac{37}{90} & \frac{79}{90} & -\frac{11}{90} & -\frac{11}{90} & \frac{19}{15} & \frac{4}{15} & -\frac{11}{15} &
   \frac{149}{30} & \frac{89}{30} & \frac{59}{30} \\
  \hline\phantom{\Big|}  45 & \frac{9}{10} & -\frac{19}{10} & \frac{224}{45} & \frac{134}{45} & \frac{44}{45} & \frac{443}{90} & \frac{263}{90} &
   \frac{83}{90} & \frac{11}{90} & \frac{11}{90} & \frac{11}{90} & \frac{7}{5} & \frac{2}{5} & -\frac{3}{5} & \frac{47}{10} &
   \frac{27}{10} & \frac{17}{10} \\
  \hline\phantom{\Big|}  46 & \frac{319}{300} & -\frac{619}{300} & \frac{1177}{300} & \frac{1177}{300} & \frac{577}{300} & \frac{307}{50} &
   \frac{107}{50} & \frac{7}{50} & \frac{151}{75} & \frac{1}{75} & \frac{1}{75} & \frac{469}{300} & \frac{169}{300} &
   -\frac{131}{300} & \frac{403}{75} & \frac{253}{75} & \frac{178}{75} \\
  \hline\phantom{\Big|}  47 & \frac{71}{60} & -\frac{131}{60} & \frac{91}{20} & \frac{71}{20} & \frac{31}{20} & \frac{169}{30} & \frac{79}{30} &
   \frac{19}{30} & \frac{19}{15} & \frac{4}{15} & \frac{4}{15} & \frac{101}{60} & \frac{41}{60} & -\frac{19}{60} &
   \frac{77}{15} & \frac{47}{15} & \frac{32}{15} \\
  \hline\phantom{\Big|}  48 & \frac{391}{300} & -\frac{691}{300} & \frac{1553}{300} & \frac{953}{300} & \frac{353}{300} & \frac{769}{150} &
   \frac{469}{150} & \frac{169}{150} & \frac{13}{25} & \frac{13}{25} & \frac{13}{25} & \frac{541}{300} & \frac{241}{300} &
   -\frac{59}{300} & \frac{367}{75} & \frac{217}{75} & \frac{142}{75}
\\ \hline
\end{array}
$
\caption{\label{48sets}The numerical results for the 48 possible 
$X$-charge assignments of Case~I, determined from 
Tables~\ref{parametersI}+\ref{Table2}.}
\end{center}\end{table}

\begin{table}
\begin{center}
\tiny
$~\hspace{-10mm}
\begin{array}{|r||r|r|c|r|c|r|c|c|c|l|}
\hline
 \# & \Delta^L_{21}  & \Delta^L_{31} & 3\zeta+p & \Delta^H & x & y & \mathrm{spectrum} & \begin{array}{c} \mathrm{maximal}\\\mathrm{denominator}\end{array} 
& \begin{array}{c}\mathrm{anomalies:}\\ k_C,\sum_i \Delta_i^{\ol{N}}    \end{array} & \mathrm{comments}  \\
\hline\hline\phantom{\Big|} 
 1 & 0 & 0 & -1 & -3 & 0 & -1 &  \text{deg.}  & 210 & 
\begin{array}{ll}
 3 & 23
\end{array}
 &  
\mathrm{\phantom{CKM}\hspace{10mm}}(\sum_i m_i \approx 5.0\,\mathrm{eV})
  \\\hline\phantom{\Big|}
 2 & 0 & 0 & -1 & -3 & 0 & 0 &  \text{deg.}  & 210 & 
\begin{array}{ll}
 3 & 23
\end{array}
 & 
\mathrm{CKM \hspace{11mm}}(\sum_i m_i \approx 5.0\,\mathrm{eV})
  \\\hline\phantom{\Big|}
 3 & 0 & 0 & -1 & -3 & 0 & 1 &  \text{deg.}  & 210 & 
\begin{array}{ll}
 3 & 23
\end{array}
 &  
\mathrm{\phantom{CKM}\hspace{10mm}}(\sum_i m_i \approx 5.0\,\mathrm{eV})

 \\\hline\phantom{\Big|}
 4 & 0 & 0 & -1 & -3 & 1 & -1 &  \text{deg.}  & 240 & 
\begin{array}{ll}
 3 & 23
\end{array}
 &  
\mathrm{\phantom{CKM}\hspace{10mm}}(\sum_i m_i \approx 3.2\,\mathrm{eV})
 \\\hline\phantom{\Big|}
 5 & 0 & 0 & -1 & -3 & 1 & 0 &  \text{deg.}  & 240 & 
\begin{array}{ll}
 3 & 23
\end{array}
 &   
\mathrm{CKM \hspace{11mm}}(\sum_i m_i \approx 3.2\,\mathrm{eV})
\\\hline\phantom{\Big|}
 6 & 0 & 0 & -1 & -3 & 1 & 1 &  \text{deg.}  & 240 & 
\begin{array}{ll}
 3 & 23
\end{array}
 &  
\mathrm{\phantom{CKM}\hspace{10mm}}(\sum_i m_i \approx 3.2\,\mathrm{eV})
 \\\hline\phantom{\Big|}
 7 & 0 & 0 & -1 & -3 & 2 & -1 &  \text{deg.}  & 270 & 
\begin{array}{ll}
 3 & 23
\end{array}
 &  
\mathrm{\phantom{CKM}\hspace{10mm}}(\sum_i m_i \approx 2.1\,\mathrm{eV})
 \\\hline\phantom{\Big|}
 8 & 0 & 0 & -1 & -3 & 2 & 0 &  \text{deg.}  & 54 & 
\begin{array}{ll}
 3 & 23
\end{array}
 & 
\mathrm{CKM,denom. \hspace{1.4mm}}(\sum_i m_i \approx 2.1\,\mathrm{eV})
  \\\hline\phantom{\Big|}
 9 & 0 & 0 & -1 & -3 & 2 & 1 &  \text{deg.}  & 270 & 
\begin{array}{ll}
 3 & 23
\end{array}
 &  
\mathrm{\phantom{CKM}\hspace{10mm}}(\sum_i m_i \approx 2.1\,\mathrm{eV})   
 \\\hline\phantom{\Big|}
 10 & 0 & 0 & -1 & -3 & 3 & -1 &  \text{deg.}  & 300 & 
\begin{array}{ll}
 3 & 23
\end{array}
 &
\mathrm{\phantom{CKM}\hspace{10mm}}(\sum_i m_i \approx 1.4\,\mathrm{eV}) 
 \\\hline\phantom{\Big|}
 11 & 0 & 0 & -1 & -3 & 3 & 0 &  \text{deg.}  & 300 & 
\begin{array}{ll}
 3 & 23
\end{array}
 & 
\mathrm{CKM \hspace{11mm}}(\sum_i m_i \approx 1.4\,\mathrm{eV})
 \\\hline\phantom{\Big|}
 12 & 0 & 0 & -1 & -3 & 3 & 1 &  \text{deg.}  & 300 & 
\begin{array}{ll}
 3 & 23
\end{array}
 & 
\mathrm{\phantom{CKM}\hspace{10mm}}(\sum_i m_i \approx 1.4\,\mathrm{eV})
  \\\hline\phantom{\Big|}
 13 & 0 & 0 & -1 & -2 & 0 & -1 & \text{inv. \& nor. hier.} & 210 & 
\begin{array}{ll}
 3 & 20
\end{array}
 &  
\mathrm{}
 \\\hline\phantom{\Big|}
 14 & 0 & 0 & -1 & -2 & 0 & 0 & \text{inv. \& nor. hier.} & 210 & 
\begin{array}{ll}
 3 & 20
\end{array}
 &  
\mathrm{CKM}
 \\\hline\phantom{\Big|}
 15 & 0 & 0 & -1 & -2 & 0 & 1 & \text{inv. \& nor. hier.} & 210 & 
\begin{array}{ll}
 3 & 20
\end{array}
 &  
\mathrm{}
 \\\hline\phantom{\Big|}
 16 & 0 & 0 & -1 & -2 & 1 & -1 & \text{inv. \& nor. hier.} & 240 & 
\begin{array}{ll}
 3 & 20
\end{array}
 & 
\mathrm{}
  \\\hline\phantom{\Big|}
 17 & 0 & 0 & -1 & -2 & 1 & 0 & \text{inv. \& nor. hier.} & 48 & 
\begin{array}{ll}
 3 & 20
\end{array}
 &  
\mathrm{CKM,denom.}
 \\\hline\phantom{\Big|}
 18 & 0 & 0 & -1 & -2 & 1 & 1 & \text{inv. \& nor. hier.} & 240 & 
\begin{array}{ll}
 3 & 20
\end{array}
 & 
\mathrm{}
  \\\hline\phantom{\Big|}
 19 & 0 & 0 & -1 & -2 & 2 & -1 & \text{inv. \& nor. hier.} & 18 & 
\begin{array}{ll}
 3 & 20
\end{array}
 &  
\mathrm{denom.}
 \\\hline\phantom{\Big|}
 20 & 0 & 0 & -1 & -2 & 2 & 0 & \text{inv. \& nor. hier.} & 90 & 
\begin{array}{ll}
 3 & 20
\end{array}
 &  
\mathrm{CKM}
 \\\hline\phantom{\Big|}
 21 & 0 & 0 & -1 & -2 & 2 & 1 & \text{inv. \& nor. hier.} & 30 & 
\begin{array}{ll}
 3 & 20
\end{array}
 &  
\mathrm{denom.}
 \\\hline\phantom{\Big|}
 22 & 0 & 0 & -1 & -2 & 3 & -1 & \text{inv. \& nor. hier.} & 12 & 
\begin{array}{ll}
 3 & 20
\end{array}
 &  
\mathrm{denom.}
 \\\hline\phantom{\Big|}
 23 & 0 & 0 & -1 & -2 & 3 & 0 & \text{inv. \& nor. hier.} & 300 & 
\begin{array}{ll}
 3 & 20
\end{array}
 & 
\mathrm{CKM}
  \\\hline\phantom{\Big|}
 24 & 0 & 0 & -1 & -2 & 3 & 1 & \text{inv. \& nor. hier.} & 300 & 
\begin{array}{ll}
 3 & 20
\end{array}
 &  
\mathrm{}
 \\\hline\phantom{\Big|}
 25 & 0 & -1 & -2 & -1 & 0 & -1 & \text{nor. hier.} & 30 & 
\begin{array}{ll}
 2 & 60 \\\hline
 3 & 18
\end{array}
 & 
\mathrm{CHOOZ,denom.},~
\begin{array}{c}
 \Delta^{\ol{N}}_{1 ,2,3}= 20 , g_X=0.0141 \\\hline 
\phantom{3}
\end{array}
  \\\hline\phantom{\Big|}
 26 & 0 & -1 & -2 & -1 & 0 & 0 & \text{nor. hier.} & 210 & 
\begin{array}{ll}
 2 & 60 \\\hline
 3 & 18
\end{array}
 &  
\mathrm{CHOOZ,CKM} ,\hspace{3mm}
\begin{array}{c}
 \Delta^{\ol{N}}_{1 ,2,3}= 20 , g_X=0.0142 \\\hline 
\phantom{3}
\end{array}
\\\hline\phantom{\Big|}
 27 & 0 & -1 & -2 & -1 & 0 & 1 & \text{nor. hier.} & 42 & 
\begin{array}{ll}
 2 & 60 \\\hline
 3 & 18
\end{array}
 & 
\mathrm{CHOOZ,denom.},~
\begin{array}{c}
 \Delta^{\ol{N}}_{1 ,2,3}= 20 , g_X=0.0141 \\\hline 
\phantom{3}
\end{array}
  \\\hline\phantom{\Big|}
 28 & 0 & -1 & -2 & -1 & 1 & -1 & \text{nor. hier.} & 240 & 
\begin{array}{ll}
 3 & 18
\end{array}
 & 
\mathrm{CHOOZ}
  \\\hline\phantom{\Big|}
 29 & 0 & -1 & -2 & -1 & 1 & 0 & \text{nor. hier.} & 240 & 
\begin{array}{ll}
 3 & 18
\end{array}
 & 
\mathrm{CHOOZ,CKM}
  \\\hline\phantom{\Big|}
 30 & 0 & -1 & -2 & -1 & 1 & 1 & \text{nor. hier.} & 240 & 
\begin{array}{ll}
 3 & 18
\end{array}
 &  
\mathrm{CHOOZ}
 \\\hline\phantom{\Big|}
 31 & 0 & -1 & -2 & -1 & 2 & -1 & \text{nor. hier.} & 18 & 
\begin{array}{ll}
 3 & 18
\end{array}
 & 
\mathrm{CHOOZ,denom.}
  \\\hline\phantom{\Big|}
 32 & 0 & -1 & -2 & -1 & 2 & 0 & \text{nor. hier.} & 30 & 
\begin{array}{ll}
 3 & 18
\end{array}
 & 
\mathrm{CHOOZ,CKM,denom.}
  \\\hline\phantom{\Big|}
 33 & 0 & -1 & -2 & -1 & 2 & 1 & \text{nor. hier.} & 90 & 
\begin{array}{ll}
 3 & 18
\end{array}
 & 
\mathrm{CHOOZ}
  \\\hline\phantom{\Big|}
 34 & 0 & -1 & -2 & -1 & 3 & -1 & \text{nor. hier.} & 300 & 
\begin{array}{ll}
 3 & 18
\end{array}
 & 
\mathrm{CHOOZ}
  \\\hline\phantom{\Big|}
 35 & 0 & -1 & -2 & -1 & 3 & 0 & \text{nor. hier.} & 300 & 
\begin{array}{ll}
 3 & 18
\end{array}
 &  
\mathrm{CHOOZ,CKM}
 \\\hline\phantom{\Big|}
 36 & 0 & -1 & -2 & -1 & 3 & 1 & \text{nor. hier.} & 300 & 
\begin{array}{ll}
 3 & 18
\end{array}
 & 
\mathrm{CHOOZ}
  \\\hline\phantom{\Big|}
 37 & -1 & -2 & -4 & 0 & 0 & -1 & \text{nor. hier.} & 210 & 
\begin{array}{ll}
 2 & 59 \\\hline
 3 & 17
\end{array}
 &  
\mathrm{CHOOZ} ,\hspace{8.3mm}
\begin{array}{c}
 \Delta^{\ol{N}}_{1 ,2}= 20, \Delta^{\ol{N}}_{3}= 19 , g_X=0.0145 \\\hline 
\phantom{3}
\end{array}
 \\\hline\phantom{\Big|}
 38 & -1 & -2 & -4 & 0 & 0 & 0 & \text{nor. hier.} & 210 & 
\begin{array}{ll}
 2 & 59 \\\hline
 3 & 17
\end{array}
 &  
\mathrm{CHOOZ,CKM} ,~
\begin{array}{c}
 \Delta^{\ol{N}}_{1 ,2}= 20, \Delta^{\ol{N}}_{3}= 19 , g_X=0.0145 \\\hline 
\phantom{3}
\end{array}
 \\\hline\phantom{\Big|}
 39 & -1 & -2 & -4 & 0 & 0 & 1 & \text{nor. hier.} & 210 & 
\begin{array}{ll}
 2 & 59 \\\hline
 3 & 17
\end{array}
 & 
\mathrm{CHOOZ} , \hspace{8.3mm}
\begin{array}{c}
 \Delta^{\ol{N}}_{1 ,2}= 20, \Delta^{\ol{N}}_{3}= 19 , g_X=0.0145 \\\hline 
\phantom{3}
\end{array}
  \\\hline\phantom{\Big|}
 40 & -1 & -2 & -4 & 0 & 1 & -1 & \text{nor. hier.} & 240 & 
\begin{array}{ll}
 3 & 17
\end{array}
 &  
\mathrm{CHOOZ}
 \\\hline\phantom{\Big|}
 41 & -1 & -2 & -4 & 0 & 1 & 0 & \text{nor. hier.} & 240 & 
\begin{array}{ll}
 3 & 17
\end{array}
 & 
\mathrm{CHOOZ,CKM}
  \\\hline\phantom{\Big|}
 42 & -1 & -2 & -4 & 0 & 1 & 1 & \text{nor. hier.} & 240 & 
\begin{array}{ll}
 3 & 17
\end{array}
 &  
\mathrm{CHOOZ}
 \\\hline\phantom{\Big|}
 43 & -1 & -2 & -4 & 0 & 2 & -1 & \text{nor. hier.} & 30 & 
\begin{array}{ll}
 3 & 17
\end{array}
 &  
\mathrm{CHOOZ,denom.}
 \\\hline\phantom{\Big|}
 44 & -1 & -2 & -4 & 0 & 2 & 0 & \text{nor. hier.} & 90 & 
\begin{array}{ll}
 3 & 17
\end{array}
 &  
\mathrm{CHOOZ,CKM}
 \\\hline\phantom{\Big|}
 45 & -1 & -2 & -4 & 0 & 2 & 1 & \text{nor. hier.} & 90 & 
\begin{array}{ll}
 3 & 17
\end{array}
 & 
\mathrm{CHOOZ}
  \\\hline\phantom{\Big|}
 46 & -1 & -2 & -4 & 0 & 3 & -1 & \text{nor. hier.} & 300 & 
\begin{array}{ll}
 3 & 17
\end{array}
 & 
\mathrm{CHOOZ}
  \\\hline\phantom{\Big|}
 47 & -1 & -2 & -4 & 0 & 3 & 0 & \text{nor. hier.} & 60 & 
\begin{array}{ll}
 3 & 17
\end{array}
 &
\mathrm{CHOOZ,CKM}
   \\\hline\phantom{\Big|}      
 48 & -1 & -2 & -4 & 0 & 3 & 1 & \text{nor. hier.} & 300 & 
\begin{array}{ll}
 3 & 17
\end{array}
 & 
\mathrm{CHOOZ}
\\\hline
\end{array}
$
\caption{\label{48details}The features of the $X$-charge assignments 
  in Table~\ref{48sets} (Case~I).
  In the comments we state the reason
  for preferring individual cases: ``CKM'' means that this model naturally
  exhibits a nice CKM matrix, {\it i.e.} $y=0$. 
  ``CHOOZ'' refers to a naturally small CHOOZ angle: $\sin{\theta_{13}}
  \approx \eps^{|\Delta^L_{31}|}$, with $|\Delta^L_{31}|=1,2$.
  We write ``denom.'' to label cases where the $X$-charges have a maximal
  denominator $\leq ~54$.
  For the degenerate scenarios we show the na\"ive sum of the neutrino masses,
  $\sum_i m_i$, without $\mathcal{O}(1)$ coefficients. Assuming no exotic
  matter, the three $\Delta_i^{\ol{N}}$ are uniquely fixed for $k_C=2$, unlike
  for $k_C=3$.}
\end{center}
\end{table}

\begin{table}
\begin{center}
\tiny
$~\hspace{-7mm}
\begin{array}{|r||r|r|r|r|r|r|r|r|r|r|r|r|r|r|r|r|r|}
\hline
\phantom{\Bigg|}\# & X_{H^D} & X_{H^U} & X_{Q^1} & X_{Q^2} & X_{Q^3} & X_{\overline{U^1}} & X_{\overline{U^2}} & X_{\overline{U^3}} & X_{\overline{D^1}} & X_{\overline{D^2}} & X_{\overline{D^3}} & 
X_{L^1} & X_{L^2} & X_{L^3} & X_{\overline{E^1}} & X_{\overline{E^2}} & X_{\overline{E^3}} \\
\hline   \hline                                                              
\phantom{\Bigg|} 1 & -  \frac{2453}{210}    & 
    \frac{2243}{210} & -  \frac{64}{35}  
      & -  \frac{64}{35}    & -  
      \frac{134}{35}    & -  \frac{179}
     {210}    & -  \frac{1019}{210}  
      & -  \frac{1439}{210}    & \frac{3677}
   {210} & \frac{3257}{210} & \frac{3257}{210} & 
    \frac{1556}{105} & \frac{1556}{105} & \frac{1556}
   {105} & \frac{391}{210} & -  \frac{239}{210}
        & -  \frac{659}{210}  
      \\\hline\phantom{\Bigg|} 2 & -  \frac{2417}{210}    & 
    \frac{2207}{210} & -  \frac{128}{105}  
      & -  \frac{233}{105}    & -  
      \frac{443}{105}    & -  \frac{271}
     {210}    & -  \frac{901}{210}  
      & -  \frac{1321}{210}    & \frac{1171}
   {70} & \frac{1101}{70} & \frac{1101}{70} & \frac{1574}
   {105} & \frac{1574}{105} & \frac{1574}{105} & 
    \frac{319}{210} & -  \frac{311}{210}  
      & -  \frac{731}{210}    \\\hline\phantom{\Bigg|} 3 & - 
     \frac{2381}{210}    & \frac{2171}{210} & 
    -  \frac{64}{105}    & -  \frac{274}
     {105}    & -  \frac{484}{105}  
      & -  \frac{121}{70}    & -  
      \frac{261}{70}    & -  \frac{401}
     {70}    & \frac{3349}{210} & \frac{3349}
   {210} & \frac{3349}{210} & \frac{1592}{105} & 
    \frac{1592}{105} & \frac{1592}{105} & \frac{247}
   {210} & -  \frac{383}{210}    & - 
     \frac{803}{210}    \\\hline\phantom{\Bigg|} 4 & -  \frac{2347}
     {210}    & \frac{2137}{210} & -  
      \frac{163}{105}    & -  \frac{163}
     {105}    & -  \frac{373}{105}  
      & -  \frac{131}{210}    & -  
      \frac{971}{210}    & -  \frac{1391}
     {210}    & \frac{1171}{70} & \frac{1031}
   {70} & \frac{1031}{70} & \frac{1504}{105} & \frac{
     1504}{105} & \frac{1399}{105} & \frac{389}
   {210} & -  \frac{241}{210}    & - 
     \frac{451}{210}    \\\hline\phantom{\Bigg|} 5 & -  \frac{2311}
     {210}    & \frac{2101}{210} & -  
      \frac{33}{35}    & -  \frac{68}{35}
        & -  \frac{138}{35}    & - 
     \frac{223}{210}    & -  \frac{853}
     {210}    & -  \frac{1273}{210}  
      & \frac{3349}{210} & \frac{3139}{210} & \frac{3139}
   {210} & \frac{1522}{105} & \frac{1522}{105} & 
    \frac{1417}{105} & \frac{317}{210} & -  
      \frac{313}{210}    & -  \frac{523}
     {210}    \\\hline\phantom{\Bigg|} 6 & -  \frac{65}{6}  
      & \frac{59}{6} & -  \frac{1}{3}  
      & -  \frac{7}{3}    & -  \frac{13}
     {3}    & -  \frac{3}{2}    & - \frac{7}{2}    & -  \frac{11}{2}
        & \frac{91}{6} & \frac{91}{6} & \frac{91}
   {6} & \frac{44}{3} & \frac{44}{3} & \frac{41}
   {3} & \frac{7}{6} & -  \frac{11}{6}  
      & -  \frac{17}{6}    \\\hline\phantom{\Bigg|} 7 & - 
     \frac{361}{30}    & \frac{331}{30} & - 
     \frac{29}{15}    & -  \frac{29}{15}
        & -  \frac{59}{15}    & - 
     \frac{11}{10}    & -  \frac{51}{10}
        & -  \frac{71}{10}    & 
    \frac{539}{30} & \frac{479}{30} & \frac{479}
   {30} & \frac{232}{15} & \frac{232}{15} & \frac{217}
   {15} & \frac{47}{30} & -  \frac{43}{30}  
      & -  \frac{73}{30}    \\\hline\phantom{\Bigg|} 8 & - 
     \frac{2491}{210}    & \frac{2281}{210} & 
    -  \frac{139}{105}    & -  \frac{244}
     {105}    & -  \frac{454}{105}  
      & -  \frac{323}{210}    & -  
      \frac{953}{210}    & -  \frac{1373}
     {210}    & \frac{1203}{70} & \frac{1133}
   {70} & \frac{1133}{70} & \frac{1642}{105} & \frac{
     1642}{105} & \frac{1537}{105} & \frac{257}
   {210} & -  \frac{373}{210}    & - 
     \frac{583}{210}    \\\hline\phantom{\Bigg|} 9 & -  \frac{491}
     {42}    & \frac{449}{42} & -  \frac{5}
     {7}    & -  \frac{19}{7}    & 
    -  \frac{33}{7}    & -  \frac{83}
     {42}    & -  \frac{167}{42}  
      & -  \frac{251}{42}    & \frac{689}
   {42} & \frac{689}{42} & \frac{689}{42} & \frac{332}
   {21} & \frac{332}{21} & \frac{311}{21} & \frac{37}
   {42} & -  \frac{89}{42}    & -  
      \frac{131}{42}    \\\hline\phantom{\Bigg|} 10 & -  \frac{
       1103}{90}    & \frac{1013}{90} & - 
     \frac{296}{135}    & -  \frac{296}
     {135}    & -  \frac{566}{135}  
      & -  \frac{287}{270}    & -  
      \frac{1367}{270}    & -  \frac{1907}
     {270}    & \frac{5521}{270} & \frac{4981}
   {270} & \frac{4981}{270} & \frac{821}{45} & \frac{821}
   {45} & \frac{776}{45} & \frac{91}{90} & -  
      \frac{179}{90}    & -  \frac{269}
     {90}    \\\hline\phantom{\Bigg|} 11 & -  \frac{1091}{90}
        & \frac{1001}{90} & -  \frac{212}
     {135}    & -  \frac{347}{135}  
      & -  \frac{617}{135}    & -  
      \frac{419}{270}    & -  \frac{1229}
     {270}    & -  \frac{1769}{270}  
      & \frac{5317}{270} & \frac{5047}{270} & \frac{5047}
   {270} & \frac{827}{45} & \frac{827}{45} & \frac{782}
   {45} & \frac{67}{90} & -  \frac{203}{90}  
      & -  \frac{293}{90}    \\\hline\phantom{\Bigg|} 12 & - 
     \frac{1079}{90}    & \frac{989}{90} & - 
     \frac{128}{135}    & -  \frac{398}
     {135}    & -  \frac{668}{135}  
      & -  \frac{551}{270}    & -  
      \frac{1091}{270}    & -  \frac{1631}
     {270}    & \frac{5113}{270} & \frac{5113}
   {270} & \frac{5113}{270} & \frac{833}{45} & \frac{833}
   {45} & \frac{788}{45} & \frac{43}{90} & -  
      \frac{227}{90}    & -  \frac{317}
     {90}    \\\hline\phantom{\Bigg|} 13 & -  \frac{2393}{210}
        & \frac{2183}{210} & -  \frac{167}
     {105}    & -  \frac{167}{105}  
      & -  \frac{377}{105}    & -  
      \frac{169}{210}    & -  \frac{1009}
     {210}    & -  \frac{1429}{210}  
      & \frac{1189}{70} & \frac{1049}{70} & \frac{1049}
   {70} & \frac{1586}{105} & \frac{1481}{105} & 
    \frac{1376}{105} & \frac{271}{210} & -  
      \frac{149}{210}    & -  \frac{359}
     {210}    \\\hline\phantom{\Bigg|} 14 & -  \frac{2357}
     {210}    & \frac{2147}{210} & -  
      \frac{103}{105}    & -  \frac{208}
     {105}    & -  \frac{418}{105}  
      & -  \frac{87}{70}    & -  
      \frac{297}{70}    & -  \frac{437}
     {70}    & \frac{3403}{210} & \frac{3193}
   {210} & \frac{3193}{210} & \frac{1604}{105} & 
    \frac{1499}{105} & \frac{1394}{105} & \frac{199}
   {210} & -  \frac{221}{210}    & - 
     \frac{431}{210}    \\\hline\phantom{\Bigg|} 15 & -  \frac{
       2321}{210}    & \frac{2111}{210} & - 
     \frac{13}{35}    & -  \frac{83}{35}
        & -  \frac{153}{35}    & - 
     \frac{353}{210}    & -  \frac{773}
     {210}    & -  \frac{1193}{210}  
      & \frac{3239}{210} & \frac{3239}{210} & \frac{3239}
   {210} & \frac{1622}{105} & \frac{1517}{105} & 
    \frac{1412}{105} & \frac{127}{210} & -  
      \frac{293}{210}    & -  \frac{503}
     {210}    \\\hline\phantom{\Bigg|} 16 & -  \frac{2573}
     {210}    & \frac{2363}{210} & -  
      \frac{69}{35}    & -  \frac{69}{35}
        & -  \frac{139}{35}    & - 
     \frac{269}{210}    & -  \frac{1109}
     {210}    & -  \frac{1529}{210}  
      & \frac{3827}{210} & \frac{3407}{210} & \frac{3407}
   {210} & \frac{1706}{105} & \frac{1601}{105} & 
    \frac{1496}{105} & \frac{211}{210} & -  
      \frac{209}{210}    & -  \frac{419}
     {210}    \\\hline\phantom{\Bigg|} 17 & -  \frac{2537}
     {210}    & \frac{2327}{210} & -  
      \frac{143}{105}    & -  \frac{248}
     {105}    & -  \frac{458}{105}  
      & -  \frac{361}{210}    & -  
      \frac{991}{210}    & -  \frac{1411}
     {210}    & \frac{1221}{70} & \frac{1151}
   {70} & \frac{1151}{70} & \frac{1724}{105} & \frac{
     1619}{105} & \frac{1514}{105} & \frac{139}
   {210} & -  \frac{281}{210}    & - 
     \frac{491}{210}    \\\hline\phantom{\Bigg|} 18 & -  \frac{
       2501}{210}    & \frac{2291}{210} & - 
     \frac{79}{105}    & -  \frac{289}
     {105}    & -  \frac{499}{105}  
      & -  \frac{151}{70}    & -  
      \frac{291}{70}    & -  \frac{431}
     {70}    & \frac{3499}{210} & \frac{3499}
   {210} & \frac{3499}{210} & \frac{1742}{105} & 
    \frac{1637}{105} & \frac{1532}{105} & \frac{67}
   {210} & -  \frac{353}{210}    & - 
     \frac{563}{210}    \\\hline\phantom{\Bigg|} 19 & -  \frac{
       2809}{240}    & \frac{2569}{240} & - 
     \frac{437}{240}    & -  \frac{437}
     {240}    & -  \frac{917}{240}  
      & -  \frac{53}{60}    & -  
      \frac{293}{60}    & -  \frac{413}
     {60}    & \frac{741}{40} & \frac{661}
   {40} & \frac{661}{40} & \frac{4031}{240} & \frac{3791}
   {240} & \frac{3551}{240} & \frac{109}{120} & - 
     \frac{131}{120}    & -  \frac{251}
     {120}    \\\hline\phantom{\Bigg|} 20 & -  \frac{2773}
     {240}    & \frac{2533}{240} & -  
      \frac{289}{240}    & -  \frac{529}
     {240}    & -  \frac{1009}{240}  
      & -  \frac{27}{20}    & -  
      \frac{87}{20}    & -  \frac{127}
     {20}    & \frac{2131}{120} & \frac{2011}
   {120} & \frac{2011}{120} & \frac{4067}{240} & 
    \frac{3827}{240} & \frac{3587}{240} & \frac{73}
   {120} & -  \frac{167}{120}    & - 
     \frac{287}{120}    \\\hline\phantom{\Bigg|} 21 & -  \frac{
       2737}{240}    & \frac{2497}{240} & - 
     \frac{47}{80}    & -  \frac{207}{80}
        & -  \frac{367}{80}    & - 
     \frac{109}{60}    & -  \frac{229}
     {60}    & -  \frac{349}{60}  
      & \frac{2039}{120} & \frac{2039}{120} & \frac{2039}
   {120} & \frac{4103}{240} & \frac{3863}{240} & 
    \frac{3623}{240} & \frac{37}{120} & -  
      \frac{203}{120}    & -  \frac{323}
     {120}    \\\hline\phantom{\Bigg|} 22 & -  \frac{1121}{90}
        & \frac{1031}{90} & -  \frac{302}
     {135}    & -  \frac{302}{135}  
      & -  \frac{572}{135}    & -  
      \frac{329}{270}    & -  \frac{1409}
     {270}    & -  \frac{1949}{270}  
      & \frac{5587}{270} & \frac{5047}{270} & \frac{5047}
   {270} & \frac{857}{45} & \frac{812}{45} & \frac{767}
   {45} & \frac{37}{90} & -  \frac{143}{90}  
      & -  \frac{233}{90}    \\\hline\phantom{\Bigg|} 23 & - 
     \frac{1109}{90}    & \frac{1019}{90} & - 
     \frac{218}{135}    & -  \frac{353}
     {135}    & -  \frac{623}{135}  
      & -  \frac{461}{270}    & -  
      \frac{1271}{270}    & -  \frac{1811}
     {270}    & \frac{5383}{270} & \frac{5113}
   {270} & \frac{5113}{270} & \frac{863}{45} & \frac{818}
   {45} & \frac{773}{45} & \frac{13}{90} & -  
      \frac{167}{90}    & -  \frac{257}
     {90}    \\\hline\phantom{\Bigg|} 24 & -  \frac{1097}{90}
        & \frac{1007}{90} & -  \frac{134}
     {135}    & -  \frac{404}{135}  
      & -  \frac{674}{135}    & -  
      \frac{593}{270}    & -  \frac{1133}
     {270}    & -  \frac{1673}{270}  
      & \frac{5179}{270} & \frac{5179}{270} & \frac{5179}
   {270} & \frac{869}{45} & \frac{824}{45} & \frac{779}
   {45} & -  \frac{11}{90}    & -  
      \frac{191}{90}    & -    \frac{281}
     {90}         \\
\hline
\end{array}
$
\caption{\label{24sets}The numerical results for the $X$-charge assignments of
  Case~II which allow no further matter to be introduced. These 24 models are
  obtained from the 504 distinct sets of Table~\ref{parametersII}.}
\end{center}\end{table}

\begin{table}
\begin{center}
\tiny
$
\begin{array}{|r||r|r|c|c|c|c|r|c|c|c|c|c|c|c|l|}
\hline
\phantom{\Bigg|} \# & \Delta^L_{21}  & \Delta^L_{31} & 3\zeta+p & \Delta^H & x & n & y & \mathrm{spectrum} & \begin{array}{c} \mathrm{max.}\\\mathrm{denom.}\end{array} 
& k_C & \Delta_1^{\ol{N}} &  \Delta_2^{\ol{N}} &    \Delta_3^{\ol{N}} &  g_X  & \mathrm{comments}    
 \\\hline\hline
\phantom{\Bigg|}  
1 & 0 & 0 & -1 & 26 & 0 & 8 & -1 &  {\mathrm{inv.~ \&~ nor.~ hier.}} & 210 & {  2} & {{ { }}  {-4}} & {{ { }}  {-9}} & {{ {}}     {-9}} & 0.0100 & 
\mathrm{}
 \\\hline\phantom{\Bigg|} 
2 & 0 & 0 & -1 & 26 & 0 & 8 & 0 &  {\mathrm{inv.~ \&~ nor.~ hier.}} & 210 & {  2} & {{ { }}  {-4}} & {{ { }}  {-9}} & 
    {{ { }}  {-9}} & 0.0100 &  
\mathrm{CKM}
  \\\hline\phantom{\Bigg|} 
3 & 0 & 0 & -1 & 26 & 0 & 8 & 1 &  {\mathrm{inv.~ \&~ nor. hier.}} & 210 & {  2} & {{ { }}  {-4}} & {{ {
       }}  {-9}} & {{ { }}  {-9}} & 0.0100 &    

\\\hline\phantom{\Bigg|} 
4 & 0 & -1 & -2 & 25 & 0 & 7 & -1 &  {\mathrm{nor.~hier.}} & 210 & {  2} &
{-2} & {{ { }}  {-8}} & {{ { }}  {-8}} & 0.0107 &   
\mathrm{CHOOZ}
 \\\hline\phantom{\Bigg|} 
5 & 0 & -1 & -2 & 25 & 0 & 7 & 0 &  {\mathrm{nor.~hier.}
   } & 210 & {  2} & {{ { }}  {-2}} & {{ { }}  {-8}} & {{ { }}  {-8}} & 0.0108
   &   
\mathrm{CHOOZ,CKM}
 \\\hline\phantom{\Bigg|} 
6 & 0 & -1 & 
    -2 & 25 & 0 & 7 & 1 &  {\mathrm{nor.~hier.}} & 6 & {  2} & {{ { }}  {-2}} & {{ { }}  {-8}} & {{ { }}  {-8}} & 
   0.0108 &   
\mathrm{CHOOZ,denom.}
 \\\hline\phantom{\Bigg|} 
7 & 0 & -1 & -2 & 27 & 0 & 8 & -1 &  {\mathrm{nor.~hier.}} & 30 & {  2} & {{ { }}  {-6}} & {{ { }}  {-9}} & {{
       { }}  {-9}} & 0.0096 &   
\mathrm{CHOOZ,denom.}
 \\\hline\phantom{\Bigg|} 
8 & 0 & -1 & -2 & 27 & 0 & 8 & 0 &  {\mathrm{nor.~hier.}} & 210 & {  2} & {{ { }}  {-6}} & {{
       { }}  {-9}} & {{ { }}  {-9}} & 0.0096 &   
\mathrm{CHOOZ,CKM}
 \\\hline\phantom{\Bigg|} 
9 & 0 & -1 & -2 & 27 & 0 & 8 & 1 &  {\mathrm{nor.~hier.}} & 42 & {  2} & {{
       { }}  {-6}} & {{ { }}  {-9}} & {{ { }}  {-9}} & 0.0096 &   
\mathrm{CHOOZ,denom.}
 \\\hline\phantom{\Bigg|} 
10 & 0 & -1 & -2 & 30 & 2 & 9 & -1 &  {\mathrm{nor.~hier.}} & 270 & {  2} & {{
    { }}  {-1}} & {{ { }}  {-10}} & {{ { }}  {-10}} & 0.0086 &    
\mathrm{CHOOZ}
\\\hline\phantom{\Bigg|} 
11 & 0 &   -1 & -2 & 30 & 2 & 9 & 0 &  {\mathrm{nor.~hier.}} & 270 & {  2} & {{ { }}  {-1}} & {{ { }}  {-10}} & {{ { }}  {-10}} & 
   0.0086 &   
\mathrm{CHOOZ,CKM}
 \\\hline\phantom{\Bigg|} 
12 & 0 & -1 & -2 & 30 & 2 & 9 & 1 &  {\mathrm{nor.~hier.}} & 270 & {  2} & {{ { }}  {-1}} & {{ { }}  {-10}} & {{
       { }}  {-10}} & 0.0086 &  
\mathrm{CHOOZ}
  \\\hline\phantom{\Bigg|} 
13 & -1 & -2 & -4 & 26 & 0 & 7 & -1 &  {\mathrm{nor.~hier.}} & 210 & {  2} &
{{ { }}  {-3}} & {  { { }}  {-8}} & {{ { }}  {-8}} & 0.0105 &   
\mathrm{CHOOZ}
 \\\hline\phantom{\Bigg|} 
14 & -1 & -2 & -4 & 26 & 0 & 7 & 0 &  {\mathrm{nor.~hier.}} & 210 & {  2} & {{
       { }}  {-3}} & {{ { }}  {-8}} & {{ { }}  {-8}} & 0.0105 &  
\mathrm{CHOOZ,CKM}
  \\\hline\phantom{\Bigg|} 
15 & -1 & -2 & -4 & 26 & 0 & 7 & 1 &  {{\mathrm{nor.~hier.}} } & 210 & {  2} &
{{ { }}  {-3}} & {{ { }}  {-8}} & {{ { }}  {-8}} & 0.0105 &
\mathrm{CHOOZ}
\\\hline\phantom{\Bigg|} 
16 & -1 & 
    -2 & -4 & 28 & 0 & 8 & -1 &  {\mathrm{nor.~hier.}} & 210 & {  2} & {{ { }}  {-7}} & {{ { }}  {-9}} & {{ { }}  {-9}} & 
   0.0094 &    
\mathrm{CHOOZ}
\\\hline\phantom{\Bigg|} 
17 & -1 & -2 & -4 & 28 & 0 & 8 & 0 &  {\mathrm{nor.~hier.}} & 210 & {  2} & {{ { }}  {-7}} & {{ { }}  {-9}} & {{
       { }}  {-9}} & 0.0094 &    
\mathrm{CHOOZ,CKM}
\\\hline\phantom{\Bigg|} 
18 & -1 & -2 & -4 & 28 & 0 & 8 & 1 &  {\mathrm{nor.~hier.}} & 210 & {  2} & {{ { }}  {-7}} & {{
       { }}  {-9}} & {{ { }}  {-9}} & 0.0094 &  
\mathrm{CHOOZ}
  \\\hline\phantom{\Bigg|} 
19 & -1 & -2 & -4 & 28 & 1 & 8 & -1 &  {\mathrm{nor.~hier.}} & 240 & {  2} & {
      { { }}  {-1}} & {{ { }}  {-9}} & {{ { }}  {-9}} & 0.0096 &   
\mathrm{CHOOZ}
 \\\hline\phantom{\Bigg|} 
20 & -1 & -2 & -4 & 28 & 1 & 8 & 0 &  
   {\mathrm{nor.~hier.}} & 240 & {  2} & {{ { }}  {-1}} & {{ { }}  {-9}} & {{
       { }}  {-9}} & 0.0097 & 
\mathrm{CHOOZ,CKM}
 \\\hline\phantom{\Bigg|} 
21 & 
    -1 & -2 & -4 & 28 & 1 & 8 & 1 &  {\mathrm{nor.~hier.} } & 240 & {  2} & {{ { }}  {-1}} & {{ { }}  {-9}} & {{ { }}  
    {-9}} & 0.0097 &   
\mathrm{CHOOZ}
 \\\hline\phantom{\Bigg|} 
22 & -1 & -2 & -4 & 31 & 2 & 9 & -1 &  {\mathrm{nor.~hier.}} & 270 & {  2} & {{ { }}  {-2}} & {{ { }}  
    {-10}} & {{ { }}  {-10}} & 0.0085 &   
\mathrm{CHOOZ}
 \\\hline\phantom{\Bigg|} 
23 & -1 & -2 & -4 & 31 & 2 & 9 & 0 &  {\mathrm{nor.~hier.}} & 270 & {  2} & {{ { }}  
    {-2}} & {{ { }}  {-10}} & {{ { }}  {-10}} & 0.0085 &  
\mathrm{CHOOZ,CKM}
  \\\hline\phantom{\Bigg|} 
24 & -1 & -2 & -4 & 31 & 2 & 9 & 1 &  {\mathrm{nor.~hier.} } & 270 & {   
    2} & {{ {  }}  {-2}} & {{ {  }}  {-10}} & {{ {   }}  {-10}} & 0.0085 &
\mathrm{CHOOZ} 
   \\\hline     
\end{array}
$
\caption{\label{24details}The features of the $X$-charge assignments in
  Table~\ref{24sets} (Case~II). In the comments we state the reason for
  preferring individual cases: ``CKM'' refers to a nice CKM matrix, ``CHOOZ''
  to a naturally small CHOOZ angle ($|\Delta^L_{31}|=1,2$), and ``denom.''
  labels models where the $X$-charges have a maximal denominator $\leq ~42$.}
\end{center}
\end{table}

\newpage
\end{appendix}

\bibliographystyle{hunsrt}
\bibliography{references}
\end{document}